\documentclass[a4paper,12pt]{article}
\pdfoutput=1
\usepackage{graphicx}
\usepackage{mathtools}
\usepackage{amssymb}
\usepackage{amsfonts}
\usepackage[footnotesize]{caption}
\usepackage[font=scriptsize]{subcaption}
\usepackage{color}
\usepackage{braket}
\usepackage{cite}
\usepackage{hyperref}
\usepackage{url}
\usepackage{multirow}
\usepackage{relsize}
\usepackage{fullpage}
\usepackage{makecell}
\usepackage{blkarray}
\usepackage{fullpage}

\newcommand{\newc}{\newcommand}
\newc{\be}{\begin{equation}}
\newc{\ee}{\end{equation}}
\newc{\bea}{\begin{eqnarray}}
\newc{\eea}{\end{eqnarray}}
\newc{\simlt}{~\mbox{\smaller\(\lesssim\)}~}
\newc{\simgt}{~\mbox{\smaller\(\gtrsim\)}~}

\begin{document}

\begin{titlepage}
\begin{center}
{\bf\Large
\boldmath{
Fermion Mass Hierarchies from Modular Symmetry}
} \\[12mm]
Simon~J.~D.~King$^{\star}$\footnote{E-mail: \texttt{sjd.king@soton.ac.uk}}, Stephen~F.~King$^{\star}$%
\footnote{E-mail: \texttt{king@soton.ac.uk}}
\\[-2mm]
\end{center}
\vspace*{0.50cm}
\centerline{$^{\star}$ \it
Department of Physics and Astronomy, University of Southampton,}
\centerline{\it
SO17 1BJ Southampton, United Kingdom }
\vspace*{1.20cm}

\begin{abstract}
{\noindent
We show how quark and lepton mass hierarchies can be reproduced in the framework of modular symmetry.
The mechanism is analogous to the Froggatt-Nielsen (FN) mechanism, but without requiring any Abelian symmetry 
to be introduced, nor any Standard Model (SM) singlet flavon to break it. The modular weights of fermion fields 
play the role of FN charges, and SM singlet fields with non-zero modular weight called weightons play the role of flavons.
We illustrate the mechanism by analysing $A_4$ (modular level 3) models of quark and lepton (including neutrino) masses and mixing, with a single modulus field. We discuss two examples in some detail, both numerically and analytically, showing how both fermion mass and mixing hierarchies emerge from different aspects of the modular symmetry.
}
\end{abstract}
\end{titlepage}

\section{\label{sec:introduction}Introduction}

The origin of the three families of quarks and leptons and their extreme range of masses remains a mystery of particle physics. According to the Standard Model (SM), quarks and leptons come in complete families that interact identically with the gauge forces, leading to a remarkably successful quantitative theory describing practically all data at the quantum level. The various quark and lepton masses are described by having different interaction strengths with the Higgs doublet, also leading to quark mixing and charge-parity (CP) violating transitions involving strange, bottom and charm quarks. However, the SM provides no understanding of the pattern of quark and lepton masses, quark mixing or CP violation.

The discovery of neutrino mass and mixing makes the flavour puzzle hard to ignore, with the fermion mass hierarchy now spanning at least 12 orders of magnitude, from the neutrino to the top quark. However, it is not only the fermion mass hierarchy that is unsettling. There are now 28 free parameters in a Majorana-extended SM, of which 22 are associated with flavour, surely too many for a fundamental theory of nature. While the quark mixing angles are small, the lepton sector has two large mixing angles $\theta_{12}$, $\theta_{23}$ and one small mixing angle $\theta_{13}$ which is of the same order of magnitude as the quark Cabibbo mixing angle~\cite{Tanabashi:2018oca}.

One early attempt to understand the quark and lepton mass hierarchies is 
the Froggatt-Nielsen (FN) mechanism \cite{Froggatt:1978nt}.
This approach assumes an additional $U(1)_{FN}$ symmetry under which the quarks and leptons carry various charges
and a cut-off scale $M_{FN}$ is associated with the breaking of the $U(1)_{FN}$ symmetry. In the SM the top quark mass of 173 GeV is given by a Yukawa coupling times the Higgs vacuum expectation value of 246 GeV divided by the square root of two. This implies a top quark Yukawa coupling close to unity. From this point of view, the top quark mass is not at all puzzling - it is the other fermion masses associated with much smaller Yukawa couplings that require explanation. According to FN, the fermions are assigned various $U(1)_{FN}$ charges and small Yukawa couplings are forbidden at the renormalisable level due to the $U(1)_{FN}$ symmetry.
The symmetry is broken by the vacuum expectation value of a new ``flavon'' field $\theta$, where $\theta$ is a neutral scalar under the SM but carries one unit of $U(1)_{FN}$  charge. Small effective Yukawa couplings then originate from non-renormalisable contact operators
where the fermion charges are compensated by powers of $\theta$, leading to suppression by powers of the small ratio 
$\langle \theta \rangle /M_{fl}$  (where $M_{fl}$ acts as a cut-off scale of the contact interaction).

To account for family replication and to address the question of large lepton mixing, theorists have explored a larger non-Abelian family 
symmetry, $SU(3)_{fl}$~\cite{King:2001uz,King:2003rf}, where the three families are analogous to the three quark colours in quantum chromodynamics (QCD). Many other examples have been proposed based on subgroups of $SU(3)_{fl}$, including non-abelian discrete flavour symmetry
(for reviews see e.g.~\cite{Altarelli:2010gt,Ishimori:2010au,King:2013eh,King:2014nza,King:2015aea,King:2017guk,Feruglio:2019ktm}).
Moreover, the leptonic CP violation phases can be predicted and the precisely measured quark CKM mixing matrix can be accommodated if the discrete flavour symmetry is combined with generalized CP symmetry~\cite{Lu:2016jit,Li:2017abz,Lu:2018oxc,Lu:2019gqp}.
However the main drawback of all such approaches that the flavour symmetry must be broken 
down to different subgroups in the neutrino and charged lepton sectors at low energy
and this requires flavon fields to obtain vacuum expectation values (VEVs) along specific directions in order to reproduce phenomenologically viable lepton mixing angles. As a consequence, the scalar potential of discrete flavour symmetry models is rather elaborate, and auxiliary abelian symmetries are usually needed to forbid dangerous operators.

Recently, modular symmetry has been suggested as the origin of flavour symmetry, with 
neutrino masses as complex analytic functions called modular forms~\cite{Feruglio:2017spp}. The starting point of this novel idea is that non-Abelian discrete family symmetries may arise from superstring theory in compactified extra dimensions, as a finite subgroup of the modular symmetry of such theories (i.e. the symmetry associated with the non-unique choice of basis vectors spanning a given extra-dimensional lattice). It follows that the 4D effective Lagrangian must respect modular symmetry. This implies that Yukawa couplings may be modular forms. So if the leptons transform as triplets under some finite modular symmetry, then the Yukawa couplings must transform nontrivially under the modular symmetry and they are modular forms which are holomorphic functions of a complex modulus field $\tau$~\cite{Feruglio:2017spp}. 
At a stroke, this removes the need for flavon fields and 
ad hoc vacuum alignments to break the family symmetry, and potentially greatly simplifies the particle content of the theory.
Moreover, all higher-dimensional operators in the superpotential are completely determined by modular invariance if supersymmetry is exact. Models with modular flavour symmetry can be highly predictive; the neutrino masses and mixing parameters can be predicted in terms of few input parameters, although the predictive power of this framework may be reduced by the K$\ddot{\mathrm{a}}$hler potential which is less  constrained by modular symmetry~\cite{Chen:2019ewa}.

The finite modular groups $\Gamma_2\cong S_3$~\cite{Kobayashi:2018vbk,Kobayashi:2018wkl,Kobayashi:2019rzp,Okada:2019xqk}, $\Gamma_3\cong A_4$~\cite{Feruglio:2017spp,Criado:2018thu,Kobayashi:2018vbk,Kobayashi:2018scp,Okada:2018yrn,Kobayashi:2018wkl,Novichkov:2018yse,Nomura:2019yft,Ding:2019zxk,Gui-JunDing:2019wap}, $\Gamma_4\cong S_4$~\cite{Penedo:2018nmg,Novichkov:2018ovf,Kobayashi:2019mna,Gui-JunDing:2019wap,Criado:2019tzk} and $\Gamma_5\cong A_5$~\cite{Novichkov:2018nkm,Ding:2019xna,Criado:2019tzk} have been considered.
For example, simple $A_4$ modular models can reproduce the measured neutrino masses and mixing angles\cite{Feruglio:2017spp,Kobayashi:2018scp,Ding:2019zxk}. 
The quark masses and mixing angles may also be included together with leptons in an $A_4$ modular invariant model~\cite{Okada:2019uoy}.
The modular invariance approach has been extended to include odd weight modular forms which can be decomposed into irreducible representations of the homogeneous finite modular group $\Gamma'_{N}$~\cite{Liu:2019khw}, and the modular symmetry 
$\Gamma'_{3}\cong T'$ has been discussed, including the new possibility of texture zeroes
\cite{Lu:2019vgm}. Also modular symmetry may be combined with generalized CP symmetry, where the modulus transforms as $\tau\rightarrow-\tau^{*}$ under the CP transformation~\cite{Novichkov:2019sqv,Baur:2019kwi,Acharya:1995ag,Dent:2001cc,Giedt:2002ns}. The formalism of the single modulus has been generalized to the case of a direct product of multiple moduli~\cite{deMedeirosVarzielas:2019cyj,King:2019vhv},
which is motivated by the additional extra dimensions in superstring theory, assuming toroidal compactification. 
Indeed, from a top-down perspective, modular symmetry naturally appears in string constructions~\cite{Kobayashi:2018rad,Kobayashi:2018bff,Baur:2019kwi,Baur:2019iai,Kobayashi:2020hoc}.

It has been realised that, if the VEV of the modulus $\tau$ takes some special value, a residual subgroup of the finite modular symmetry group $\Gamma_N$ would be preserved. The phenomenological implications of the residual modular symmetry have been discussed in the context of modular $A_4$~\cite{Novichkov:2018yse,Gui-JunDing:2019wap}, $S_4$~\cite{Gui-JunDing:2019wap,Novichkov:2018ovf} and $A_5$~\cite{Novichkov:2018nkm} symmetries. 
If the modular symmetry is broken down to a residual $Z_3$ (or $Z_5$) subgroup in charged lepton sector and to a $Z_2$ subgroup in the neutrino sector, the trimaximal TM1 and TM2 mixing patterns can be obtained~\cite{Novichkov:2018yse,Novichkov:2018ovf}.

In this paper, we show how fermion mass hierarchies can be reproduced in the framework of modular symmetry.
The mechanism is analogous to the FN mechanism, but without requiring any Abelian symmetry 
to be introduced, nor any SM singlet flavon to break it. The modular weights of fermion fields 
play the role of FN charges, and a SM singlet field $\phi$ with non-zero modular weight (called a ``weighton'') plays the role of a flavon. We illustrate the mechanism with modular level 3 ($A_4$) models of quark and lepton (including neutrino) masses
and mixing, using a single modulus field $\tau$ and where the charged fermion mass hierarchies originate from a single weighton $\phi$. We discuss two such viable models in some detail, both numerically and analytically, showing how both fermion mass and mixing hierarchies emerge from the modular symmetry.
The class of modular level 3 (with even weight modular forms) examples
of the mechanism we present here is by no means exhaustive; the new mechanism may be applied to other levels and choices of weights, and to models with any number of moduli fields and weightons.

We highlight the two of the new features of this work compared to previous analyses.
\begin{itemize}
\item{We introduce a new mechanism for charged fermion mass hierarchies analogous to the FN mechanism, but without any flavons,
using instead modular weights and a complete singlet ``weighton'' field $\phi$. The only other paper to address the charged lepton mass hierarchy
with modular symmetry is \cite{Criado:2019tzk} which obtained suppression for charged lepton
masses in a level 4 ($S_4$) model using a triplet flavon, whereas here our weighton is a complete singlet. }
\item{We discuss quark mass hierarchies using our mechanism.
There are only two papers which address the quark masses and mixings using modular symmetry \cite{Okada:2019uoy,Lu:2019vgm},
and neither paper has attempted to account for the quark mass hierarchies. In \cite{Lu:2019vgm}, texture zeroes in the quark Yukawa matrices
were enforced using odd weight modular forms, in particular the double cover of $A_4$, namely $\Gamma'_{3}\cong T'$, but no attempt was made to account for the quark or charged lepton mass hierarchies.
}
\end{itemize}

We also remark that the approach here differs from an early work based on $U(1)_{FN}$ broken by a flavon $\theta$, where all fields carried both FN charge and modular weight \cite{Leontaris:1997vw}. In our approach, the Yukawa couplings are modular forms, which means that the modular weights do not have to sum to zero, and are triplets under the $A_4$ modular symmetry,
which constrains the rows of the Yukawa matrices. We also emphasise that we do not have any $U(1)_{FN}$ symmetry, nor any flavon $\theta$ to break such a symmetry. Our weighton $\phi$ is an $A_4$ singlet which does not break any flavour symmetry and is therefore not a flavon.

The layout of the remainder of the paper is as follows. In section~\ref{sec:modularform_of_N=4} 
we briefly review modular symmetry.
In section~\ref{level3} we give the necessary results for level 3 modular forms. 
In section~\ref{model} we show how a previously proposed $A_4$ modular model of leptons can be recast in natural form by introducing a single weighton, then apply similar ideas to 27 possible models in the quark sector. In section~\ref{results} we analyse all the quark models, combined with the natural lepton model, and 
identify two viable combinations,
which can successfully describe all quark and lepton (including neutrino) masses and mixing, using a single modulus field $\tau$, and in which all charged fermion mass hierarchies originate from a single weighton. 
We discuss these two viable models in some detail, both numerically and analytically, showing how all fermion mass and mixing hierarchies emerge from the modular symmetry.
Section~\ref{conclusion} concludes the paper.

\section{\label{sec:modularform_of_N=4}Modular symmetry }

The modular group $\overline{\Gamma}$ is the group of linear fraction transformations which acts on the complex modulus $\tau$ in the upper half complex plane as follow,
\begin{equation}
\tau\rightarrow\gamma\tau=\frac{a\tau+b}{c\tau+d},~~\text{with}~~a, b, c, d\in\mathbb{Z},~~ad-bc=1,~~~\Im\tau>0\,.
\end{equation}
We note that the map
\begin{equation}
\frac{a\tau+b}{c\tau+d}\mapsto\left(
\begin{array}{cc}
a  &  b  \\
c  &  d
\end{array}
\right)
\end{equation}
is an isomorphism from the modular group to the projective
matrix group $PSL(2, \mathbb{Z})\cong SL(2, \mathbb{Z})/\{\pm I\} $, where $SL(2, \mathbb{Z})$ is the group of two-by-two matrices with integer entries and determinant equal to one. 

The modular group $\overline{\Gamma}$ can be generated by two generators $S$ and $T$
\begin{equation}
S:\tau\mapsto -\frac{1}{\tau},~~~~\quad T: \tau\mapsto\tau+1\,,
\end{equation}
which are represented by the following two matrices of $PSL(2, \mathbb{Z})$,
\begin{equation}
S=\left(\begin{array}{cc}
0 & 1 \\
-1  & 0
\end{array}
\right),~~~\quad T=\left(\begin{array}{cc}
1 & 1 \\
0  & 1
\end{array}
\right)\,.
\end{equation}
We can check that the generators $S$ and $T$ obey the relations,
\begin{equation}
\label{eq:multiply_rules}S^2=(ST)^3=(TS)^3=1\,.
\end{equation}
The principal congruence subgroup of level $N$ is the subgroup
\begin{equation}
\Gamma(N)=\left\{\begin{pmatrix}
a  & b \\
c  & d
\end{pmatrix}\in SL(2, \mathbb{Z}),~ b=c=0\,(\mathrm{mod~N}), a=d=1\,(\mathrm{mod~N})
\right\}\,,
\end{equation}
which is an infinite normal subgroup of $SL(2, \mathbb{Z})$. It is easy to see that $T^{N}$ is an element of $\Gamma(N)$. The projective principal congruence subgroup is defined as $\overline{\Gamma}(N)=\Gamma(N)/\{\pm I \}$ for $N=1, 2$. For the values of $N\geq3$, we have $\overline{\Gamma}(N)=\Gamma(N)$ because
$\Gamma(N)$ doesn't contain the element $-I$. The quotient group $\Gamma_N\equiv\overline{\Gamma}/\overline{\Gamma}(N)$ is the finite modular group, and it can be obtained by further imposing the condition $T^{N}=1$ besides those in Eq.~\eqref{eq:multiply_rules}.

A crucial element of the modular invariance approach is the modular form $f(\tau)$ of weight $k$ and level $N$. The modular form $f(\tau)$ is a holomorphic function of the complex modulus $\tau$ and it is required to transform under the action of $\overline{\Gamma}(N)$ as follows,
\begin{equation}
f\left(\frac{a\tau+b}{c\tau+d}\right)=(c\tau+d)^kf(\tau)~~~\mathrm{for}~~\forall~\left(
\begin{array}{cc}
a  &  b  \\
c  &  d
\end{array}
\right)\in\overline{\Gamma}(N)\,.
\end{equation}
The modular forms of weight $k$ and level $N$ span a linear space of finite dimension. It is always possible to choose a basis in this linear space such that the modular forms can be arranged into some modular multiplets $f_{\mathbf{r}}\equiv\left(f_1(\tau), f_{2}(\tau),...\right)^{T}$ which transform as irreducible representation $\mathbf{r}$ of the finite modular group $\Gamma_N$ for even $k$~\cite{Feruglio:2017spp,Liu:2019khw}, i.e.
\begin{equation}
f_{\mathbf{r}}(\gamma\tau)=(c\tau+d)^k\rho_{\mathbf{r}}(\gamma)f_{\mathbf{r}}(\tau)~~~\mathrm{for}~~\forall~\gamma\in\overline{\Gamma}\,,
\end{equation}
where $\gamma$ is the representative element of the coset
$\gamma\overline{\Gamma}(N)$ in $\Gamma_N$, and $\rho_{\mathbf{r}}(\gamma)$ is the representation matrix of the element $\gamma$ in the irreducible representation $\mathbf{r}$.

The superpotential $W(\Phi_I,\tau)$ can be expanded in power series of the supermultiplets $\Phi_I$,
\begin{equation}
W(\Phi_I,\tau) =\sum_n Y_{I_1...I_n}(\tau)~ \Phi_{I_1}... \Phi_{I_n}\,,
\end{equation}
where $Y_{I_1...I_n}$ is a modular multiplet of weight $k_Y$ and it transforms in the representation $\rho_{Y}$ of $\Gamma_{N}$,
\begin{equation}
\begin{aligned}
&\tau\to \gamma\tau =\dfrac{a\tau+b}{c\tau+d}\,,\\
&Y(\tau)\to Y(\gamma\tau)=(c\tau+d)^{k_Y}\rho_{Y}(\gamma)Y(\tau)\,.
\end{aligned}
\end{equation}
The requirement of modular invariance of the superpotential implies
\begin{equation}
k_Y=k_{I_1}+...+k_{I_n},~\quad~ \rho_Y\otimes\rho_{I_1}\otimes\ldots\otimes\rho_{I_n}\ni\mathbf{1}\,.
\end{equation}
where the supermultiplet $\Phi_{I_1}$ is assumed to transform in a representation $\rho_{I_1}$ of $\Gamma_{N}$,
with a modular weight $-k_{I_1}$, and so on for the other supermultiplets.

\section{Modular forms of $\Gamma_3\cong A_{4}$ (level 3)}
\label{level3}

The modular group $\Gamma(3)$ has been extensively studied in the literature~\cite{Feruglio:2017spp,Criado:2018thu,Kobayashi:2018vbk,Kobayashi:2018scp,Okada:2018yrn,Kobayashi:2018wkl,Novichkov:2018yse,Gui-JunDing:2019wap,Nomura:2019yft}. In the present work we shall adopt the same convention as~\cite{Feruglio:2017spp,Gui-JunDing:2019wap,Ding:2019zxk}. The finite modular group $\Gamma_3$ is isomorphic to $A_4$ which is the symmetry group of the tetrahedron. It contains twelve elements and it is the smallest non-abelian finite group which admits an irreducible three-dimensional representation. The $A_4$ group has three one-dimensional representations $\mathbf{1}$, $\mathbf{1}'$, $\mathbf{1}''$ and a three-dimensional representation $\mathbf{3}$. In the singlet representations, we have
\begin{equation}
\begin{aligned}
& \mathbf{1}:~~ S=1, \qquad T=1 \,,  \\
& \mathbf{1}^{\prime}:~~ S=1, \qquad T=\omega^{2} \,,  \\
&\mathbf{1}^{\prime\prime}:~~S=1, \qquad T=\omega \,.
\end{aligned}
\end{equation}
For the representation $\mathbf{3}$, we will choose a basis in which the generator $T$ is diagonal. The explicit forms of $S$ and $T$ are
\begin{equation}
S=\frac{1}{3}\begin{pmatrix}
    -1& 2  & 2  \\
    2  & -1  & 2 \\
    2 & 2 & -1
\end{pmatrix}, ~\quad~
T=\begin{pmatrix}
    1 ~&~ 0 ~&~ 0 \\
    0 ~&~ \omega^{2} ~&~ 0 \\
    0 ~&~ 0 ~&~ \omega
\end{pmatrix} \,,
\end{equation}
with $\omega=e^{2\pi i/3}=-1/2+i\sqrt{3}/2$. The basic multiplication rule is
\begin{equation}
\mathbf{3}\otimes \mathbf{3}= \mathbf{1}\oplus \mathbf{1'}\oplus \mathbf{1''}\oplus \mathbf{3}_S\oplus \mathbf{3}_A\,,
\end{equation}
where the subscripts $S$ and $A$ denotes symmetric and antisymmetric combinations respectively. If we have two triplets $\alpha=(\alpha_1,\alpha_2,\alpha_3)\sim\mathbf{3}$ and  $\beta=(\beta_1,\beta_2,\beta_3)\sim\mathbf{3}$, we can obtain the following irreducible representations from their product,
\begin{eqnarray}
\nonumber &&(\alpha\beta)_{\mathbf{1}}=\alpha_1\beta_1+\alpha_2\beta_3+\alpha_3\beta_2\,, \\
\nonumber &&(\alpha\beta)_{\mathbf{1}^{\prime}}=\alpha_3\beta_3+\alpha_1\beta_2+\alpha_2\beta_1\,, \\
\nonumber &&(\alpha\beta)_{\mathbf{1}^{\prime\prime}}=\alpha_2\beta_2+\alpha_1\beta_3+\alpha_3\beta_1\,, \\
\nonumber &&(\alpha\beta)_{\mathbf{3}_S}=(
2\alpha_1\beta_1-\alpha_2\beta_3-\alpha_3\beta_2,
2\alpha_3\beta_3-\alpha_1\beta_2-\alpha_2\beta_1,
2\alpha_2\beta_2-\alpha_1\beta_3-\alpha_3\beta_1)\,, \\
\label{eq:CG_coefficient} &&(\alpha\beta)_{\mathbf{3}_A}=(
\alpha_2\beta_3-\alpha_3\beta_2,
\alpha_1\beta_2-\alpha_2\beta_1,
\alpha_3\beta_1-\alpha_1\beta_3)\,.
\end{eqnarray}
The linear space of the modular forms of integral weight $k$ and level $N=3$  has dimension $k+1$~\cite{Feruglio:2017spp}. The modular space $\mathcal{M}_{2k}(\Gamma(3))$ can be constructed from the Dedekind eta-function $\eta(\tau)$ which is defined as
\begin{equation}
\eta(\tau)=q^{1/24} \prod_{n =1}^\infty (1-q^n), \qquad  q=e^{2\pi i\tau}\,.
\end{equation}
The Dedekind eta-function $\eta(\tau)$ satisfies the following identities
\begin{equation}
\eta(\tau+1)=e^{i \pi/12}\eta(\tau),\qquad \eta(-1/\tau)=\sqrt{-i \tau}~\eta(\tau)\,.
\end{equation}
There are only three linearly independent modular forms of weight 2 and level 3, which are denoted as $Y_i(\tau)$ with $i=1, 2, 3$. We can arrange the three modular functions into a vector $Y^{(2)}_{\mathbf{3}}=\left(Y_1, Y_2, Y_3\right)^{T}$ transforming as a triplet $\mathbf{3}$ of $A_4$. The modular forms $Y_i$ can be expressed in terms of $\eta(\tau)$ and its derivative as follow~\cite{Feruglio:2017spp}:
\begin{eqnarray}
Y_1(\tau) &=& \frac{i}{2\pi}\left[ \frac{\eta'(\tau/3)}{\eta(\tau/3)}  +\frac{\eta'((\tau +1)/3)}{\eta((\tau+1)/3)}
+\frac{\eta'((\tau +2)/3)}{\eta((\tau+2)/3)} - \frac{27\eta'(3\tau)}{\eta(3\tau)}  \right], \nonumber \\
Y_2(\tau) &=& \frac{-i}{\pi}\left[ \frac{\eta'(\tau/3)}{\eta(\tau/3)}  +\omega^2\frac{\eta'((\tau +1)/3)}{\eta((\tau+1)/3)}
+\omega \frac{\eta'((\tau +2)/3)}{\eta((\tau+2)/3)}  \right] ,\nonumber \\
Y_3(\tau) &=& \frac{-i}{\pi}\left[ \frac{\eta'(\tau/3)}{\eta(\tau/3)}  +\omega\frac{\eta'((\tau +1)/3)}{\eta((\tau+1)/3)}
+\omega^2 \frac{\eta'((\tau +2)/3)}{\eta((\tau+2)/3)} \right]\,.
\end{eqnarray}
The $q$-expansions of the triplet modular forms $Y^{(2)}_{\mathbf{3}}$ are given by
\begin{eqnarray}
Y^{(2)}_{\mathbf{3}}=\begin{pmatrix}Y_1(\tau)\\Y_2(\tau)\\Y_3(\tau)\end{pmatrix}=
\begin{pmatrix}
1 + 12q + 36q^2 + 12q^3 + 84q^4 + 72q^5 +\dots \\
-6q^{1/3}(1 + 7q + 8q^2 + 18q^3 + 14q^4 +\dots) \\
-18q^{2/3}(1 + 2q + 5q^2 + 4q^3 + 8q^4 +\dots)
\end{pmatrix}\,.
\label{Y2}
\end{eqnarray}
They satisfy the constraint~\cite{Feruglio:2017spp}
\begin{equation}
(Y^{(2)}_{\mathbf{3}}Y^{(2)}_{\mathbf{3}})_{\mathbf{1}''}\equiv Y_2^2+2 Y_1 Y_3=0\,.
\end{equation}
Multiplets of higher weight modular forms can be constructed from the tensor products of $Y^{(2)}_{\mathbf{3}}$. 

Using the $A_4$ contraction $\mathbf{3}\otimes \mathbf{3}= \mathbf{1}\oplus \mathbf{1'}\oplus \mathbf{1''}\oplus \mathbf{3}_S\oplus \mathbf{3}_A$, we can obtain five independent weight 4 modular forms,
\begin{equation}
\begin{aligned}
&Y^{(4)}_{\mathbf{1}}=Y_1^2+2 Y_2 Y_3\sim\mathbf{1},\\
&Y^{(4)}_{\mathbf{1}'}=Y_3^2+2 Y_1 Y_2\sim\mathbf{1}'\,,\\
&Y^{(4)}_{\mathbf{3}}=\begin{pmatrix}Y^{(4)}_1\\Y^{(4)}_2\\Y^{(4)}_3\end{pmatrix}
=
\left(\begin{array}{c}
Y_1^2-Y_2 Y_3\\
Y_3^2-Y_1 Y_2\\
Y_2^2-Y_1 Y_3
\end{array}
\right)\sim\mathbf{3}\,.
\end{aligned}
\label{weight4}
\end{equation}

Similarly there are seven modular forms of weight 6, which can be decomposed as $\mathbf{1}\oplus\mathbf{3}\oplus\mathbf{3}$ under $A_4$~\cite{Feruglio:2017spp},
\begin{equation}
\begin{aligned}
&Y^{(6)}_{\mathbf{1}}=Y_1^3+Y_2^3+Y_3^3-3 Y_1 Y_2 Y_3\sim\mathbf{1}\,,\\
&Y^{(6)}_{\mathbf{3}, I}
=\begin{pmatrix}Y^{(6)}_{1,I}\\Y^{(6)}_{2,I}\\Y^{(6)}_{3,I}\end{pmatrix}
=\left(\begin{array}{c}
Y_1^3+2 Y_1Y_2Y_3\\
Y_1^2Y_2+2Y_2^2Y_3 \\
Y_1^2Y_3+2Y_3^2Y_2
\end{array}\right)\,,\\
&Y^{(6)}_{\mathbf{3}, II}
=\begin{pmatrix}Y^{(6)}_{1,II}\\Y^{(6)}_{2,II}\\Y^{(6)}_{3,II}\end{pmatrix}
=\left(\begin{array}{c}
Y_3^3+2 Y_1Y_2Y_3\\
Y_3^2Y_1+2Y_1^2Y_2\\
Y_3^2Y_2+2Y_2^2Y_1
\end{array}\right)\,.
\end{aligned}
\label{weight6}
\end{equation}

It has been realised that, if the VEV of the modulus $\tau$ takes some special value, a residual subgroup of the finite modular symmetry group 
$\Gamma_3$ would be preserved. Thus, the fixed points $\tau_S=i$, $\tau_{ST}=(-1+i\sqrt{3})/2$, $\tau_{TS}=(1+i\sqrt{3})/2$, $\tau_T=i\infty$ in the fundamental domain are invariant under modular transformations, and there are many other examples in the upper half complex plane~\cite{Gui-JunDing:2019wap}. 
For example, $\tau_T=i\infty$ implies $Y^{(2)}_{\mathbf{3}}\propto \left(1, 0, 0\right)^{T}$,
$Y^{(4)}_{\mathbf{3}}\propto \left(1, 0, 0\right)^{T}$, $Y^{(6)}_{\mathbf{3}, I}\propto \left(1, 0, 0\right)^{T}$, 
$Y^{(6)}_{\mathbf{3}, II}\propto \left(0, 0, 0\right)^{T}$.

\section{Models with $\Gamma_3\cong A_{4}$ (level 3)}
\label{model}

\subsection{The Feruglio model of leptons}

\begin{table}[t!]
\renewcommand{\tabcolsep}{0.5mm}
\begin{center}
\begin{tabular}{|c|c|c|c|c|c|c|c|}\hline\hline
& $L$ & $e^c_3$  &  $e^c_2$ &  $e^c_1 $ & $N^{c}$  &$H_{u,d}$   \\ \hline

$A_4$  & $\mathbf{3}$ &  $\mathbf{1}'$  &  $\mathbf{1}''$ &  $\mathbf{1}$ & $\mathbf{3}$  & $\mathbf{1}$     
\\ \hline

$k_I$ & $1$ & $1$  & $1$ & $1$  & $1$ &  $0$    \\ \hline\hline

\end{tabular}
\caption{\label{tab:model1} The Feruglio model of leptons, where each supermultiplet has a modular weight $-k_I$.}
\end{center}
\end{table}

In this subsection we review an example of a model of lepton masses and mixing based on $A_4$ modular symmetry, first introduced as example 3 in~\cite{Feruglio:2017spp} and later reanalysed in the light of current data in~\cite{Ding:2019zxk}.
In this example, there is no flavon field other than the modulus $\tau$. The Higgs doublets $H_u$ and $H_d$ are assumed to transform as $\mathbf{1}$ under $A_4$ and their modular weights $k_{H_u, H_d}$ are vanishing. The neutrino masses are assumed arise from the type I seesaw mechanism. In this example~\cite{Feruglio:2017spp}, the three generations of left-handed lepton doublets $L\equiv(L_1, L_2, L_3)^{T}$ and of the 
CP conjugated right-handed neutrino $N^c\equiv(N^c_1, N^c_2, N^c_3)^{T}$ are organised into two triplets $\mathbf{3}$ of $A_4$ with modular weights denoted as $k_L$ and $k_N$, which will be fixed to take the values of unity shown in Table~\ref{tab:model1}.

When the three CP conjugated right-handed charged leptons $e^{c}_{3,2,1}$ are assigned to
three different singlets $\mathbf{1}'$, $\mathbf{1}''$ and $\mathbf{1}$ of $A_4$ as in previous works~\cite{Feruglio:2017spp,Criado:2018thu,Kobayashi:2018vbk,Kobayashi:2018scp,Okada:2018yrn,Kobayashi:2018wkl,Novichkov:2018yse}, their modular weights could be identical, which will be fixed to take the values of unity as shown in Table~\ref{tab:model1},
and only the lowest weight modular form $Y^{(2)}_{\mathbf{3}}$ is necessary in the minimal model. 

Then the superpotential for the charged lepton masses takes the form
\begin{align}
\nonumber
W_e
&=\alpha e^c_1(LY^{(2)}_{\mathbf{3}})_{\mathbf{1}}H_d
+\beta e^c_2(LY^{(2)}_{\mathbf{3}})_{\mathbf{1}'}H_d + \gamma e^c_3(LY^{(2)}_{\mathbf{3}})_{\mathbf{1}''}H_d\\
\nonumber
&=\alpha e^c_1(L_1 Y_1+L_2 Y_3+L_3 Y_2)H_d+\beta e^c_2(L_3 Y_3+L_1 Y_2+L_2 Y_1)H_d\\
&\quad+\gamma e^c_3(L_2 Y_2+L_3 Y_1+L_1 Y_3)H_d\,.
\label{eq:We_10}
\end{align}

The invariance of $W_e$ under modular transformations implies the
following relations for the weights,
\begin{equation}
\begin{cases}
k_{e_1}+k_L=2\,,\\
k_{e_2}+k_L=2\,, \\
k_{e_3}+k_L=2\,,
\end{cases}
\label{weightconstr1}
\end{equation}
which implies
\begin{equation}
k_{e_1}=k_{e_2}=k_{e_3}=2-k_L\,,
\end{equation}
where all values are fixed to be unity as shown in Table~\ref{tab:model1}.
This is exactly the case considered in the literature~\cite{Feruglio:2017spp,Criado:2018thu,Kobayashi:2018vbk,Kobayashi:2018scp,Okada:2018yrn,Kobayashi:2018wkl,Novichkov:2018yse}. We can straightforwardly read out the charged lepton Yukawa matrix 
\begin{equation}
 Y_e =\begin{pmatrix}
	 ~~&~~     ~~&~~    ~~&~~ \\[-0.1in]
 \alpha\,Y_1 ~~&~~ \alpha\,Y_3 ~~&~~ \alpha\,Y_2 \\
 ~~&~~     ~~&~~    ~~&~~ \\[-0.02in]
\beta Y_2 ~&~ \beta Y_1  ~&~ \beta Y_3 \\
 ~&~        ~&~      ~&~  \\[-0.02in]
 \gamma Y_3~&~ \gamma Y_2~&~\gamma Y_1\\
 ~~&~~     ~~&~~    ~~&~~ \\[-0.1in]
 \end{pmatrix}  
 \label{yuke1}
 \end{equation}
 For example, $\tau_T=i\infty$ implies $Y^{(2)}_{\mathbf{3}}\propto \left(1, 0, 0\right)^{T}$, leads to a diagonal charged lepton Yukawa matrix with 
 $m_e:m_{\mu}:m_{\tau}= \alpha :\beta :\gamma $.
 The charged lepton mass hierarchies are accounted for in the Feruglio model by tuning the parameters to be 
 $\alpha \ll \beta \ll \gamma $.

 If neutrino masses are generated through the type-I seesaw mechanism, for the triplet assignments of both right-handed neutrinos $N^{c}$ and left-handed lepton doublets $L$, the most general form of the superpotential in the neutrino sector is
\begin{equation}
W_\nu = g \left(N^c L H_uf_N\left(Y\right)\right)_\mathbf{1}
+ \Lambda \left(N^c N^cf_M\left(Y\right)\right)_\mathbf{1}\,,
\label{eq:WnuII}
\end{equation}
where $f_N(Y)$ and $f_M(Y)$ are generic functions of the modular forms $Y(\tau)$. Motivated by the principle of minimality, we consider the following example:
$f_N\left(Y\right)\propto Y^{(2)}_{\mathbf{3}}$ and $f_M\left(Y\right)\propto Y^{(2)}_{\mathbf{3}}$, which implies,
\begin{align}
\nonumber
W_\nu
&=g_1((N^c\,L)_{\mathbf{3}_S}Y^{(2)}_{\mathbf{3}})_\mathbf{1}H_u+g_2((N^c\,L)_{\mathbf{3}_A}Y^{(2)}_{\mathbf{3}})_\mathbf{1}H_u
+\Lambda (\left(N^c N^c\right)_\mathbf{3_S}Y^{(2)}_{\mathbf{3}})_\mathbf{1}\\
\nonumber
&=g_1\big[(2N^c_1L_1-N^c_2L_3-N^c_3L_2)Y_1+(2N^c_3L_3-N^c_1L_2-N^c_2L_1)Y_3 \\
\nonumber
&~~+(2N^c_2L_2-N^c_3L_1-N^c_1L_3)Y_2\big]H_u+g_2\big[(N^c_2L_3-N^c_3L_2)Y_1+(N^c_1L_2-N^c_2L_1)Y_3\\
\nonumber&~~+(N^c_3L_1-N^c_1L_3)Y_2\big]H_u+2\Lambda\big[(N^c_1N^c_1-N^c_2N^c_3)Y_1+(N^c_3N^c_3-N^c_1N^c_2)Y_3 \\
&~~+(N^c_2N^c_2-N^c_1N^c_3)Y_2\big]\,.
\label{eq:Wnu_s3}
\end{align}
The modular weights of $N^c$ and $L$ correspond to $k_L=k_N=1$ as shown in Table~\ref{tab:model1}.

We find $M_D$ and $M_N$ take the following form
\begin{eqnarray}
\nonumber&&\qquad\qquad~~ M_N = \begin{pmatrix}
2Y_1 ~&~ -Y_3 ~&~ -Y_2 \\
 -Y_3 ~&~ 2Y_2  ~&~ -Y_1  \\
 -Y_2 ~&~ -Y_1 ~&~2Y_3
\end{pmatrix}\Lambda\,,\\
&&
M_D =\begin{pmatrix}
2g_1Y_1        ~&~  (-g_1+g_2)Y_3 ~&~ (-g_1-g_2)Y_2 \\
(-g_1-g_2)Y_3  ~&~     2g_1Y_2    ~&~ (-g_1+g_2)Y_1  \\
 (-g_1+g_2)Y_2 ~&~ (-g_1-g_2)Y_1  ~&~ 2g_1Y_3
\end{pmatrix}v_{u}\,.
\label{seesawmatrices}
\end{eqnarray}
The light neutrino mass matrix is given by the seesaw formula,
\begin{equation}
\label{eq:mnufactor}M_\nu\,=\,-M_D^TM^{-1}_NM_D\,.
\end{equation}
This is the original Feruglio model introduced as example 3 in~\cite{Feruglio:2017spp}, corresponding to the case of
$\mathcal{D}_{10}$ in~\cite{Ding:2019zxk}, giving an excellent fit to current experimental data.
The best fit (allowed range) of the modulus for $\mathcal{D}_{10}$ in \cite{Ding:2019zxk} is: 
$\rm{Re}\,\langle \tau \rangle= 0.0386 (0.0307 \sim 0.1175)$, $\rm{Im}\,\langle \tau \rangle= 2.230 (1.996 \sim 2.50 )$,
which approximates the fixed point case $\tau_T=i\infty$,
since the real part is much less than the imaginary part.

\subsection{A natural model of charged leptons}
In this subsection we show how Feruglio's $A_4$ modular model of charged leptons can be recast in natural form by introducing a single weighton. The neutrino sector will remain unchanged to leading order.
The resulting model of leptons shown in Table~\ref{tab:model2}
now involves a single ``weighton'' $\phi$ which is defined to be a SM and $A_4$ singlet field with $k_{\phi}=1$ (i.e. weight $-1$).
We show how such a model can 
generate a natural charged lepton mass hierarchy. In the next subsection we extend the idea to the quark sector, thereby explaining all charged fermion masses naturally.

The three right-handed charged leptons $e^{c}_{3,2,1}$ are assigned to
three different singlets $\mathbf{1}'$, $\mathbf{1}''$ and $\mathbf{1}$ of $A_4$ as before but now their modular weights are not identical, and correspond to $k_{e^{c}_{3,2,1}}=0,-1,-3$
(i.e. weights $0,1,3$)
such that powers of $\phi$ with $k_{\phi}=1$ are required compensate the terms in the 
previous model, with the combinations $e^{c}_{3}\phi, e^{c}_{2}\phi^2, e^{c}_{1}\phi^4$ each having combined weights of unity as before.
The weighton $\phi$ is assumed to develop a vacuum expectation value (vev) so that the corresponding terms are suppressed by powers of 
\begin{equation}
\tilde{\phi}\equiv \frac{\langle \phi \rangle}{M_{fl}}, 
\label{phitilde}
\end{equation}
where $M_{fl}$ is a dimensionful cut-off flavour scale.

The weighton vev in Eq.\ref{phitilde} may be driven by a 
leading order superpotential term
\begin{equation}
W_{driv}= \chi(Y^{(4)}_{\mathbf{1}}\frac{\phi^4}{M_{fl}^2} -M^2), 
\label{driving}
\end{equation}
where 
$\chi$ is an $A_4$ singlet driving superfield with zero modular weight,   
while $M$ is a free dimensionful mass scale. This is similar to the usual driving field mechanism familiar from flavon models~\cite{Altarelli:2010gt,Ishimori:2010au,King:2013eh,King:2014nza,King:2015aea,King:2017guk,Feruglio:2019ktm}, except for the presence of the lowest weight singlet modular form $Y^{(4)}_{\mathbf{1}}$ listed in Eq.\ref{weight4},
where the quadratic term $\phi^2$ is forbidden since $Y^{(2)}_{\mathbf{1}}$
does not exist, and we have dropped higher powers such as $\phi^6$, and so on. As usual~\cite{Altarelli:2010gt,Ishimori:2010au,King:2013eh,King:2014nza,King:2015aea,King:2017guk,Feruglio:2019ktm}, the structure of the driving superpotential
$W_{driv}$ may be enforced by a $U(1)_R$ symmetry, 
with the driving superfield 
$\chi$ having $R=2$, the weighton $\phi$ and Higgs superfields having $R=0$ and the matter superfields having $R=1$, which prevents other superpotential terms appearing
\footnote{At the low energy scale, after the inclusion of SUSY breaking effects, the $U(1)_R$ symmetry will be broken to the usual discrete R-parity~\cite{Altarelli:2010gt}. Such SUSY breaking effects may also modify the predictions from modular symmetry~\cite{Feruglio:2017spp}.
However the study of SUSY breaking is beyond the scope of this paper.}. 
The F-flatness condition
$F_{\chi}=\frac{\partial W_{driv}}{\partial \chi}=0$ applied to Eq.\ref{driving}
then drives a weighton vev, $\langle \phi \rangle \sim (M M_{fl})^{1/2}$, leading to the suppression factor
in Eq.\ref{phitilde} being given by $\tilde{\phi}\sim (M/M_{fl})^{1/2}$,
where we assume $M\ll M_{fl}$.

The suppression factor in Eq.\ref{phitilde} generates the charged lepton mass hierarchy naturally,  with 
$m_{\tau,\mu, e }\propto \tilde{\phi},  \tilde{\phi}^2,  \tilde{\phi}^4$, 
with only the lowest weight modular form $Y^{(2)}_{\mathbf{3}}$ being necessary as before.

\begin{table}[t!]
\renewcommand{\tabcolsep}{0.5mm}
\begin{center}
\begin{tabular}{|c|c|c|c|c|c|c|c|c|}\hline\hline
& $L$ & $e^c_3$  &  $e^c_2$ &  $e^c_1 $ & $N^{c}$  &$H_{u,d}$ &   $\phi$   
\\ \hline

$A_4$  & $\mathbf{3}$ &  $\mathbf{1}'$  &  $\mathbf{1}''$ &  $\mathbf{1}$ & $\mathbf{3}$  & $\mathbf{1}$  &  $\mathbf{1}$   
\\ \hline

$k_I$ & $1$ & $0$  & $-1$ & $-3$  & $1$ &  $0$ &$1$   \\ \hline\hline

\end{tabular}
\caption{\label{tab:model2} A natural $A_4$ model of leptons with a weighton $\phi$.
Note that each supermultiplet has a modular weight $-k_I$. }
\end{center}
\end{table}

After the weighton develops its vev, the superpotential for the charged lepton masses takes the form
\begin{align}
\nonumber
W_e
&=\alpha_e e^c_1  \tilde{\phi}^4(LY^{(2)}_{\mathbf{3}})_{\mathbf{1}}H_d
+\beta_e e^c_2 \tilde{\phi}^2(LY^{(2)}_{\mathbf{3}})_{\mathbf{1}'}H_d + \gamma_e e^c_3 \tilde{\phi}(LY^{(2)}_{\mathbf{3}})_{\mathbf{1}''}H_d\\
\nonumber
&=\alpha_e e^c_1 \tilde{\phi}^4(L_1 Y_1+L_2 Y_3+L_3 Y_2)H_d+\beta_e e^c_2 \tilde{\phi}^2(L_3 Y_3+L_1 Y_2+L_2 Y_1)H_d\\
&\quad+\gamma_e e^c_3 \tilde{\phi} (L_2 Y_2+L_3 Y_1+L_1 Y_3)H_d\,,
\label{eq:We_101}
\end{align}
which gives a charged lepton Yukawa matrix similar to Eq.\ref{yuke1}, except that it involves powers of $ \tilde{\phi}$ controlling the hierarchies,
\begin{equation}
Y_e =\begin{pmatrix}
	 ~~&~~     ~~&~~    ~~&~~ \\[-0.1in]
 \alpha_e  \tilde{\phi}^4\,Y_1 ~~&~~ \alpha_e  \tilde{\phi}^4 \,Y_3 ~~&~~ \alpha_e  \tilde{\phi}^4 \,Y_2 \\
 ~~&~~     ~~&~~    ~~&~~ \\[-0.02in]
\beta_e  \tilde{\phi}^2 Y_2 ~&~ \beta_e  \tilde{\phi}^2 Y_1  ~&~ \beta_e  \tilde{\phi}^2 Y_3 \\
 ~&~        ~&~      ~&~  \\[-0.02in]
 \gamma_e  \tilde{\phi} Y_3~&~ \gamma_e  \tilde{\phi} Y_2~&~\gamma_e  \tilde{\phi} Y_1\\
 ~~&~~     ~~&~~    ~~&~~ 
 \label{yuke2}
 \end{pmatrix} 
 \end{equation}
 For example, $\tau_T=i\infty$ implies $Y^{(2)}_{\mathbf{3}}\propto \left(1, 0, 0\right)^{T}$, leading to a diagonal 
 and naturally hierarchical charged lepton Yukawa matrix with 
 $m_e:m_{\mu}:m_{\tau}= \alpha_e \tilde{\phi}^4:\beta_e \tilde{\phi}^2:\gamma_e \tilde{\phi} $.
 The empirically observed charged lepton mass ratios $m_e/m_{\mu}=1/207$
 and $m_{\mu}/m_{\tau}=1/17$ suggest that we fix $\tilde{\phi}\approx 1/15$ to account for the charged lepton mass hierarchy,
 with the mass ratios $m_e/m_{\mu}\sim \tilde{\phi}^2$ and $m_{\mu}/m_{\tau}\sim  \tilde{\phi}$,
 assuming order one coefficients $\alpha_e, \beta_e, \gamma_e \sim 1$. The small parameter $\tilde{\phi}\approx 1/15$
defined to be the ratio of scales in Eq.\ref{phitilde} now provides an explanation for the charged lepton mass hierarchies.

However now there will be additional terms corresponding to higher weight modular forms, $Y^{(4)}_{\mathbf{3}}$,
compensated by extra powers of weighton
fields $\phi$, which will give corrections to the charged lepton superpotential,
\begin{align}
\nonumber
\Delta W_e
&=\alpha'_e e^c_1 \tilde{\phi}^6(LY^{(4)}_{\mathbf{3}})_{\mathbf{1}}H_d
+\beta'_e e^c_2 \tilde{\phi}^4(LY^{(4)}_{\mathbf{3}})_{\mathbf{1}'}H_d + \gamma'_e e^c_3 \tilde{\phi}^3(LY^{(4)}_{\mathbf{3}})_{\mathbf{1}''}H_d\\
\nonumber
&=\alpha'_e e^c_1 \tilde{\phi}^6(L_1 Y^{(4)}_1+L_2 Y^{(4)}_3+L_3 Y^{(4)}_2)H_d
+\beta'_e e^c_2 \tilde{\phi}^4(L_3 Y^{(4)}_3+L_1 Y^{(4)}_2+L_2 Y^{(4)}_1)H_d\\
&\quad+\gamma'_e e^c_3 \tilde{\phi}^3 (L_2 Y^{(4)}_2+L_3 Y^{(4)}_1+L_1 Y^{(4)}_3)H_d\,,
\end{align}
where from Eq.\ref{weight4} the weight 4 Yukawa couplings are given in terms of the weight 2 Yukawa couplings,
 \begin{equation}
Y^{(4)}_1=Y_1^2-Y_2 Y_3, \ \ \ \ Y^{(4)}_2=Y_3^2-Y_1 Y_2, \ \ \ \ Y^{(4)}_3=Y_2^2-Y_1 Y_3.
\label{Yp}
\end{equation}
This yields the additive correction to the charged lepton mass matrix in Eq.\ref{yuke2},
\begin{equation}
\Delta Y_e =\begin{pmatrix}
	 ~~&~~     ~~&~~    ~~&~~ \\[-0.1in]
 \alpha'_e  \tilde{\phi}^6\,Y^{(4)}_1 ~~&~~ \alpha'_e  \tilde{\phi}^6 \,Y^{(4)}_3 ~~&~~ \alpha'_e  \tilde{\phi}^6 \,Y^{(4)}_2 \\
 ~~&~~     ~~&~~    ~~&~~ \\[-0.02in]
\beta'_e  \tilde{\phi}^4 Y^{(4)}_2 ~&~ \beta'_e  \tilde{\phi}^4 Y^{(4)}_1  ~&~ \beta'_e  \tilde{\phi}^4 Y^{(4)}_3 \\
 ~&~        ~&~      ~&~  \\[-0.02in]
 \gamma'_e  \tilde{\phi}^3 Y^{(4)}_3~&~ \gamma'_e  \tilde{\phi}^3 Y^{(4)}_2~&~\gamma'_e  \tilde{\phi}^3 Y^{(4)}_1\\
 ~~&~~     ~~&~~    ~~&~~ 
 \label{yuke3}
 \end{pmatrix} 
 \end{equation}
 where $\alpha'_e, \beta'_e, \gamma'_e$ are new free complex coefficients (also assumed to be of order unity) while the 
 weight 4 Yukawa couplings are given in Eq.\ref{Yp}.
For example, $\tau_T=i\infty$ implies $Y^{(2)}_{\mathbf{3}}\propto \left(1, 0, 0\right)^{T}$, implies that the higher order corrections
also take the form of a diagonal charged lepton Yukawa matrix.
However these are just the leading corrections.
There will also be further corrections from even higher weight modular forms, such as $Y^{(6)}_{\mathbf{3}}$,
compensated by extra powers of weighton
fields $\phi$, which will give further corrections to the charged lepton
Yukawa matrix. However, since $\tilde{\phi}\approx 1/15$,
we find all such corrections to be very suppressed, and have a negligible effect on the numerical results.

Since the modular weights of $L$ and $N^c$ are unchanged, and their representations are the same, we expect the seesaw neutrino matrices to be the same as in the original model at lowest order, where no weighton field $\phi$ appears and $f_N\left(Y\right)\propto Y^{(2)}_{\mathbf{3}}$ and $f_M\left(Y\right)\propto Y^{(2)}_{\mathbf{3}}$ as in Eq.\ref{eq:Wnu_s3}. Thus the seesaw matrices in this model are exactly the same as
in Eq.\ref{seesawmatrices}.
However now there will higher order corrections involving weightons, the leading correction being suppressed by $\tilde{\phi}^2$,
\begin{align}
\nonumber
\Delta W_\nu
&=g'_1 \tilde{\phi}^2((N^c\,L)_{\mathbf{3}_S}Y^{(4)}_{\mathbf{3}})_\mathbf{1}H_u+g'_2\tilde{\phi}^2((N^c\,L)_{\mathbf{3}_A}Y^{(4)}_{\mathbf{3}})_\mathbf{1}H_u
+\Lambda'  \tilde{\phi}^2 (\left(N^c N^c\right)_\mathbf{3_S}Y^{(4)}_{\mathbf{3}})_\mathbf{1}\\
\nonumber
&=g'_1 \tilde{\phi}^2\big[(2N^c_1L_1-N^c_2L_3-N^c_3L_2)Y^{(4)}_1+(2N^c_3L_3-N^c_1L_2-N^c_2L_1)Y^{(4)}_3 \\
\nonumber
&~~+(2N^c_2L_2-N^c_3L_1-N^c_1L_3)Y^{(4)}_2\big]H_u
\\
\nonumber&~~+g'_2 \tilde{\phi}^2\big[(N^c_2L_3-N^c_3L_2)Y^{(4)}_1+(N^c_1L_2-N^c_2L_1)Y^{(4)}_3+(N^c_3L_1-N^c_1L_3)Y^{(4)}_2\big]H_u
 \\
 &~~+2\Lambda'  \tilde{\phi}^2\big[(N^c_1N^c_1-N^c_2N^c_3)Y^{(4)}_1+(N^c_3N^c_3-N^c_1N^c_2)Y^{(4)}_3+(N^c_2N^c_2-N^c_1N^c_3)Y^{(4)}_2\big]\,,
\label{eq:Wnu_s4p}
\end{align}
which is of the same form as in Eq.\ref{eq:Wnu_s3},
yielding additive corrections to the 
seesaw matrices of the same form as in Eq.\ref{seesawmatrices} but suppressed by $\tilde{\phi}^2$ and with 
the primed Yukawa couplings given by Eq.\ref{Yp}.
As before, since $\tilde{\phi}\approx 1/15$, these corrections are expected to be about $0.5\%$, so in the neutrino sector we can safely ignore these 
corrections and use the same results as before. Thus we expect that the modulus 
best fit to point to be the same value quoted as before, approximating the fixed point case $\tau_T=i\infty$.

\subsection{Natural models of quarks}

\begin{table}[t!]
\renewcommand{\tabcolsep}{0.5mm}
\begin{center}
\begin{tabular}{|c|c|c|c|c|c|c|c|c|c|c|}\hline\hline
& $Q$ & $d^c_3$  &  $d^c_2$ &  $d^c_1 $ & $u^c_3$  &  $u^c_2$ &  $u^c_1 $ &$H_{u,d}$ &   $\phi$    \\ \hline

$A_4$  & $\mathbf{3}$ &  $\mathbf{1}'$  &  $\mathbf{1}''$ &  $\mathbf{1}$   &  $\mathbf{1}'$  &  $\mathbf{1}''$ &  $\mathbf{1}$   &$\mathbf{1}$  &  $\mathbf{1}$   
\\ \hline

$k_I$ & $1$ & $0,2,4$  & $-2$ & $-3$  &  $5,3,1$  & $-1,2,4$ & $-3$  & $0$ &$1$    \\ \hline\hline

\end{tabular}
\caption{\label{tab:model3} Natural $A_4$ models of quarks with a weighton $\phi$.
All 27 combinations of modular weights are considered in the text.
Note that each supermultiplet has a modular weight $-k_I$. }
\end{center}
\end{table}

Quarks have been considered with $A_4$ modular symmetry in \cite{Okada:2019uoy}.
However there has been no attempt to explain the quark mass hierarchy.
Using similar ideas developed in the previous section for the charged leptons, 
we now consider models for the down type quark Yukawa matrix with 
$m_d:m_s:m_b\sim  \tilde{\phi}^4: \tilde{\phi}^3:  \tilde{\phi} $,
which turns out to be a good description of the down quark mass hierarchies as we shall see.
As in the charged lepton sector,
the weighton is assumed to develop a vacuum expectation value (vev) so that the corresponding terms are suppressed by powers of 
$ \tilde{\phi}=\langle \phi \rangle/M_{fl}$, where $M_{fl}$ is a dimensionful cut-off flavour scale,
which we assume to be the same scale as for the charged leptons.

We introduce the quark modular weights in Table~\ref{tab:model3} which can achieve this, using the same
weighton $\phi$ as in the charged lepton sector.
We assign the quark doublets $Q$ to a triplet of $A_4$ with $k_Q=1$ analogous to the lepton doublets.
The three right-handed down type quarks $d^{c}_{3,2,1}$ are assigned to
three different singlets $\mathbf{1}'$, $\mathbf{1}''$ and $\mathbf{1}$ of $A_4$,
analogous to how the charged lepton Yukawa matrix was constructed.

Unlike in the charged lepton sector, here we allow higher weight modular forms in the quark sector,
which will prove necessary to describe quark mixing.
We therefore have more freedom in assigning various modular weights to $d^{c}_{3,2,1}$
such that powers of $\phi$ with $k_{\phi}=1$ are required compensate the terms, 
with the combinations $d^{c}_{3}\phi, d^{c}_{2}\phi^3, d^{c}_{1}\phi^4$ appearing,
analogous to the charged lepton assignments. 
This generates the down type quark mass hierarchy naturally,  with 
$m_{b,s, d }\propto \tilde{\phi},  \tilde{\phi}^3,  \tilde{\phi}^4$.

After the weighton develops its VEV, the superpotential for the down type quark masses 
with $k_{d^{c}_{3,2,1}}=0,-2,-3$
takes the form
\begin{align}
\nonumber
W_d
&=\alpha_d d^c_1  \tilde{\phi}^4(QY^{(2)}_{\mathbf{3}})_{\mathbf{1}}H_d
+\beta_d d^c_2 \tilde{\phi}^3(QY^{(2)}_{\mathbf{3}})_{\mathbf{1}'}H_d + \gamma_d d^c_3 \tilde{\phi}(QY^{(2)}_{\mathbf{3}})_{\mathbf{1}''}H_d\\
\nonumber
&=\alpha_d d^c_1 \tilde{\phi}^4(Q_1 Y_1+Q_2 Y_3+Q_3 Y_2)H_d+\beta_d d^c_2 \tilde{\phi}^3(Q_3 Y_3+Q_1 Y_2+Q_2 Y_1)H_d\\
&\quad+\gamma_d d^c_3 \tilde{\phi} (Q_2 Y_2+Q_3 Y_1+Q_1 Y_3)H_d\,,
\label{eq:Wd_101}
\end{align}
which gives a similar form of Yukawa matrix for the down type quarks as for the charged leptons in Eq.\ref{yuke2}, 
albeit the second row being more suppressed than before,
\begin{equation}
Y_d^{I} =\begin{pmatrix}
	 ~~&~~     ~~&~~    ~~&~~ \\[-0.1in]
 \alpha_d  \tilde{\phi}^4\,Y_1 ~~&~~ \alpha_d  \tilde{\phi}^4 \,Y_3 ~~&~~ \alpha_d  \tilde{\phi}^4 \,Y_2 \\
 ~~&~~     ~~&~~    ~~&~~ \\[-0.02in]
\beta_d  \tilde{\phi}^3 Y_2 ~&~ \beta_d  \tilde{\phi}^3 Y_1  ~&~ \beta_d  \tilde{\phi}^3 Y_3 \\
 ~&~        ~&~      ~&~  \\[-0.02in]
 \gamma_d  \tilde{\phi} Y_3~&~ \gamma_d  \tilde{\phi} Y_2~&~\gamma_d  \tilde{\phi} Y_1\\
 ~~&~~     ~~&~~    ~~&~~ 
 \label{yukd2}
 \end{pmatrix} 
 \end{equation}
 where without loss of generality we may take $ \alpha_d , \beta_d , \gamma_u$ to be real.
However now there will be additional terms corresponding to higher weight modular forms, $Y^{(4)}_{\mathbf{3}}$,
compensated by extra powers of weighton
fields $\phi$, which will give corrections to the down type quark superpotential, analogous to the higher order corrections
to the charged lepton superpotential in Eq.\ref{eq:We_101}. Since these corrections will yield a matrix with a similar structure to 
the lowest order matrix but with each element having an additional correction 
be suppressed by a relative power of 
$\tilde{\phi}^2$.
This yields the additive correction to the down type quark mass matrix in Eq.\ref{yukd2},
\begin{equation}
\Delta Y_d =\begin{pmatrix}
	 ~~&~~     ~~&~~    ~~&~~ \\[-0.1in]
 \alpha'_d  \tilde{\phi}^6\,Y^{(4)}_1 ~~&~~ \alpha'_d  \tilde{\phi}^6 \,Y^{(4)}_3 ~~&~~ \alpha'_d  \tilde{\phi}^6 \,Y^{(4)}_2 \\
 ~~&~~     ~~&~~    ~~&~~ \\[-0.02in]
\beta'_d  \tilde{\phi}^5 Y^{(4)}_2 ~&~ \beta'_d  \tilde{\phi}^5 Y^{(4)}_1  ~&~ \beta'_d  \tilde{\phi}^5 Y^{(4)}_3 \\
 ~&~        ~&~      ~&~  \\[-0.02in]
 \gamma'_d  \tilde{\phi}^3 Y^{(4)}_3~&~ \gamma'_d  \tilde{\phi}^3 Y^{(4)}_2~&~\gamma'_d  \tilde{\phi}^3 Y^{(4)}_1\\
 ~~&~~     ~~&~~    ~~&~~ 
 \label{yukd2.1}
 \end{pmatrix} 
 \end{equation}
 where $\alpha'_d, \beta'_d, \gamma'_d$ are new free complex coefficients (also assumed to be of order unity) while the 
 weight 4 Yukawa couplings are given in Eq.\ref{Yp}.

 Other alternatives include $k_{d^{c}_{3,2,1}}=2,-2,-3$:
 \begin{equation}
Y_d^{II} =\begin{pmatrix}
	 ~~&~~     ~~&~~    ~~&~~ \\[-0.1in]
 \alpha_d  \tilde{\phi}^4\,Y_1 ~~&~~ \alpha_d  \tilde{\phi}^4 \,Y_3 ~~&~~ \alpha_d  \tilde{\phi}^4 \,Y_2 \\
 ~~&~~     ~~&~~    ~~&~~ \\[-0.02in]
\beta_d  \tilde{\phi}^3 Y_2 ~&~ \beta_d  \tilde{\phi}^3 Y_1  ~&~ \beta_d  \tilde{\phi}^3 Y_3 \\
 ~&~        ~&~      ~&~  \\[-0.02in]
 \gamma_d \tilde{\phi} Y^{(4)}_3 ~&~  \gamma_d \tilde{\phi} Y^{(4)}_2  ~&~ \gamma_d  \tilde{\phi} Y^{(4)}_1 \\
 ~~&~~     ~~&~~    ~~&~~ 
 \label{yukd2.2}
 \end{pmatrix} 
 \end{equation}
 
 Also we consider $k_{d^{c}_{3,2,1}}=4,-2,-3$:
 \begin{equation}
Y_d^{III} =\begin{pmatrix}
	 ~~&~~     ~~&~~    ~~&~~ \\[-0.1in]
 \alpha_d  \tilde{\phi}^4\,Y_1 ~~&~~ \alpha_d  \tilde{\phi}^4 \,Y_3 ~~&~~ \alpha_d  \tilde{\phi}^4 \,Y_2 \\
 ~~&~~     ~~&~~    ~~&~~ \\[-0.02in]
\beta_d  \tilde{\phi}^3 Y_2 ~&~ \beta_d  \tilde{\phi}^3 Y_1  ~&~ \beta_d  \tilde{\phi}^3 Y_3 \\
 ~&~        ~&~      ~&~  \\[-0.02in]
 \gamma^I_d \tilde{\phi} Y^{(6)}_{3,I} + \gamma^{II}_d\tilde{\phi}  Y^{(6)}_{3,II}~&~ 
 \gamma^I_d \tilde{\phi}  Y^{(6)}_{2,I}+ \gamma^{II}_d \tilde{\phi}  Y^{(6)}_{2,II}~&~
 \gamma^I_d\tilde{\phi}   Y^{(6)}_{1,I}+\gamma^{II}_d \tilde{\phi}  Y^{(6)}_{1,II}\\
  ~~&~~     ~~&~~    ~~&~~ 
 \label{yukd2.3}
 \end{pmatrix} 
 \end{equation}
 
It is worth noting that we have achieved a single power of suppression $ \tilde{\phi}$ for the third down type family in several ways,
by choosing an even weight for $k_{d^{c}_{3}}=0,2,4,\ldots$ 
so that $d^{c}_{3}\tilde{\phi}Q$ is also even and may be compensated by a Yukawa coupling modular form of weight $2,4,6,\ldots$, leading to the three possibilities for the third row of the down type Yukawa matrix as above (with more possibilities at even higher weight).
On the other hand higher powers of suppression such as $ \tilde{\phi}^3,  \tilde{\phi}^4$ for the first two families may only be achieved
at lowest order by a Yukawa coupling modular form of weight $2$.

Turning to the up type quark sector, we first consider
$m_u:m_c:m_t\sim  \tilde{\phi}^4: \tilde{\phi}^2:  1$, assuming $\tilde{\phi}\approx 1/15$ as before.
In order to achieve this, the up type quarks are assigned the modular weights as 
shown in Table~\ref{tab:model3}.
The three right-handed up type quarks $u^{c}_{3,2,1}$ are assigned to
three different singlets $\mathbf{1}'$, $\mathbf{1}''$ and $\mathbf{1}$ of $A_4$.
In this case the up type quark mass hierarchy is much stronger and the top quark Yukawa coupling is of order unity,
which suggests that it should be unsuppressed without any weighton field being involved.
Moreover, as shown in \cite{Okada:2019uoy}, the lowest weight modular forms $Y^{(2)}_{\mathbf{3}}$ 
are not sufficient to describe quark mixing so here we shall utilise weight 6
modular form $Y^{(6)}_{\mathbf{3}}$ for only the third family (whereas in \cite{Okada:2019uoy} weight 6 modular forms
were assumed for all three families of quarks).
If we had used the lowest weight modular forms $Y^{(2)}_{\mathbf{3}}$ for all three families
then the up quark Yukawa matrix would have rows proportional to that of the down quark Yukawa matrix, leading to zero quark mixing angles, so we need to use higher weight modular forms for the up Yukawa matrix, at least for the second or third families, and here we use weight 6
only for the third family.
This motivates the assignments $k_{u^{c}_{3,2,1}}=5,-1,-3$
such that the combinations $Qu^{c}_{3}, Qu^{c}_{2}\phi^2, Qu^{c}_{1}\phi^4$ 
imply the modular forms $Y^{(6)}_{\mathbf{3}}$, $Y^{(2)}_{\mathbf{3}}$, $Y^{(2)}_{\mathbf{3}}$, respectively,
where powers of $\phi$ with $k_{\phi}=1$ are required. 
Actually there are two independent weight $6$ modular forms $Y^{(6)}_{\mathbf{3},I}$ and $Y^{(6)}_{\mathbf{3},II}$
and both must be considered as contributing independently.

Although the above assignments satisfies our requirements, we need to check that these are indeed the leading order terms.
Firstly $Qu^{c}_{3}$ has weight $-6$ so the leading term is $Y^{(6)}_{\mathbf{3}}$, with the higher order correction 
$Qu^{c}_{3}\phi^2$ having weight $-8$ and requiring $Y^{(8)}_{\mathbf{3}}$ (the lower weight modular forms
$Y^{(2)}_{\mathbf{3}}$ and $Y^{(4)}_{\mathbf{3}}$ are forbidden at all orders).
Secondly, although $Qu^{c}_{2}$ has weight zero, this term 
is forbidden since it is an $A_4$ triplet and 
$Y^{(0)}_{\mathbf{3}}$ does not exist.
Therefore the leading allowed term is $Qu^{c}_{2}\phi^2$ with weight $-2$, 
compensated by $Y^{(2)}_{\mathbf{3}}$,
with the higher order term $Qu^{c}_{2}\phi^4$ with weight $-4$ compensated by $Y^{(4)}_{\mathbf{3}}$ being suppressed.
Thirdly $Qu^{c}_{1}$ has weight $2$ and cannot be compensated by a modular form with positive weight.
While $Qu^{c}_{1}\phi^2$ has weight zero it is forbidden since it is an $A_4$ triplet and triplet modular forms cannot have zero weight.
Therefore the leading term is $Qu^{c}_{1}\phi^4$ with weight $-2$ which is compensated by 
$Y^{(2)}_{\mathbf{3}}$, with the higher order correction 
$Qu^{c}_{1}\phi^6$ having weight $-4$ compensated by $Y^{(4)}_{\mathbf{3}}$ being suppressed.

After the weighton develops its vev, the leading order superpotential for the up type quark masses takes the form
\begin{align}
\nonumber
W_u
&=\alpha_u u^c_1  \tilde{\phi}^4(QY^{(2)}_{\mathbf{3}})_{\mathbf{1}}H_u
+ \beta_u u^c_2 \tilde{\phi}^2 (QY^{(2)}_{\mathbf{3}})_{\mathbf{1}'}H_u
+\gamma^I_u u^c_3(QY^{(6)}_{\mathbf{3},I})_{\mathbf{1}''}H_u 
+\gamma^{II}_u u^c_3(QY^{(6)}_{\mathbf{3},II})_{\mathbf{1}''}H_u 
\\
\nonumber
&=\alpha_u u^c_1 \tilde{\phi}^4(Q_1 Y_1+Q_2 Y_3+Q_3 Y_2)H_u
+\beta_u u^c_2 \tilde{\phi}^2(Q_3 Y_3+Q_1 Y_2+Q_2 Y_1)H_u
\\ \nonumber
&+\gamma^I_u u^c_3  (Q_2 Y^{(6)}_{2,I}+Q_3 Y^{(6)}_{1,I}+Q_1 Y^{(6)}_{3,I})H_u
+\gamma^{II}_u u^c_3  (Q_2 Y^{(6)}_{2,II}+Q_3 Y^{(6)}_{1,II}+Q_1 Y^{(6)}_{3,II})H_u\,,
\label{eq:Wu_101}
\end{align}
Thus $k_{u^{c}_{3,2,1}}=5,-1,-3$ leads to the up quark Yukawa matrix,
\begin{equation}
Y_u^{I} =\begin{pmatrix}
	 ~~&~~     ~~&~~    ~~&~~ \\[-0.1in]
 \alpha_u  \tilde{\phi}^4\,Y_1 ~~&~~ \alpha_u  \tilde{\phi}^4 \,Y_3 ~~&~~ \alpha_u  \tilde{\phi}^4 \,Y_2 \\
 ~~&~~     ~~&~~    ~~&~~ \\[-0.02in]
\beta_u  \tilde{\phi}^2 Y_2 ~&~ \beta_u  \tilde{\phi}^2 Y_1  ~&~ \beta_u  \tilde{\phi}^2 Y_3 \\
 ~&~        ~&~      ~&~  \\[-0.02in]
 \gamma^I_u  Y^{(6)}_{3,I} + \gamma^{II}_u  Y^{(6)}_{3,II}~&~ 
 \gamma^I_u   Y^{(6)}_{2,I}+ \gamma^{II}_u   Y^{(6)}_{2,II}~&~
 \gamma^I_u   Y^{(6)}_{1,I}+\gamma^{II}_u   Y^{(6)}_{1,II}\\
 ~~&~~     ~~&~~    ~~&~~ 
 \label{yuku2}
 \end{pmatrix} 
 \end{equation}
 where the weight 6 Yukawa couplings are given in Eq.\ref{weight6}.
 This is consistent with a diagonal and naturally hierarchical up type quark Yukawa matrix with 
 $m_u:m_c:m_t\sim  \tilde{\phi}^4: \tilde{\phi}^2:  1$,
 where without loss of generality we may take $ \alpha_u , \beta_u,  \gamma^I_u $ to be real,
 while in general $\gamma^{II}_u$ can be complex.

Before performing a numerical study of this case,
we recall that, $\tau_T=i\infty$ implies $Y^{(2)}_{\mathbf{3}}\propto \left(1, 0, 0\right)^{T}$,
$Y^{(4)}_{\mathbf{3}}\propto \left(1, 0, 0\right)^{T}$, $Y^{(6)}_{\mathbf{3}, I}\propto \left(1, 0, 0\right)^{T}$, 
$Y^{(6)}_{\mathbf{3}, II}\propto \left(0, 0, 0\right)^{T}$ so near this limit $Y^{(6)}_{\mathbf{3}, II}$ will not contribute.
However we need to go away from this limit to explain quark mixing angles.
There is a potential problem with the Yukawa structures in Eqs.\ref{yukd2},\ref{yuku2}
since analytically (ignoring third family mixing angles) we expect
$\theta_{12}^d\sim \theta_{12}^u \sim Y_2/Y_1$, so the physical Cabibbo angle 
$\theta_{12}\sim \theta_{12}^d - \theta_{12}^d \sim 0$ due to cancellation.

To avoid this problem we also consider an alternative model with 
the assignments $k_{u^{c}_{3,2,1}}=5,2,-3$
(i.e. only differing by the assignment $k_{u^c_2}=2$)
such that the combinations $Qu^{c}_{3}, Qu^{c}_{2}\phi, Qu^{c}_{1}\phi^4$ 
imply the modular forms $Y^{(6)}_{\mathbf{3}}$, $Y^{(4)}_{\mathbf{3}}$, $Y^{(2)}_{\mathbf{3}}$, respectively,
where powers of $\phi$ with $k_{\phi}=1$ are required. This may avoid the cancellation problem of the Cabibbo angle,
since now $\theta_{12}^u \sim Y^{(4)}_2/Y^{(4)}_1$ is different from $\theta_{12}^d\sim Y_2/Y_1$,
but is slightly less natural, being consistent with a diagonal and naturally hierarchical up type quark Yukawa matrix with 
 $m_u:m_c:m_t\sim  \tilde{\phi}^4: \tilde{\phi}:  1$. Thus $k_{u^{c}_{3,2,1}}=5,2,-3$ leads to:
 \begin{equation}
Y_u^{II} =\begin{pmatrix}
	 ~~&~~     ~~&~~    ~~&~~ \\[-0.1in]
 \alpha_u  \tilde{\phi}^4\,Y_1 ~~&~~ \alpha_u  \tilde{\phi}^4 \,Y_3 ~~&~~ \alpha_u  \tilde{\phi}^4 \,Y_2 \\
 ~~&~~     ~~&~~    ~~&~~ \\[-0.02in]
\beta_u  \tilde{\phi} Y^{(4)}_2 ~&~ \beta_u  \tilde{\phi} Y^{(4)}_1  ~&~ \beta_u  \tilde{\phi} Y^{(4)}_3 \\
 ~&~        ~&~      ~&~  \\[-0.02in]
 \gamma^I_u  Y^{(6)}_{3,I} + \gamma^{II}_u  Y^{(6)}_{3,II}~&~ 
 \gamma^I_u   Y^{(6)}_{2,I}+ \gamma^{II}_u   Y^{(6)}_{2,II}~&~
 \gamma^I_u   Y^{(6)}_{1,I}+\gamma^{II}_u   Y^{(6)}_{1,II}\\
 ~~&~~     ~~&~~    ~~&~~ 
 \label{yuku3}
 \end{pmatrix} 
 \end{equation}
The analysis of the alternative model using the up quark Yukawa matrix 
in Eq.\ref{yuku3} is very similar to that using  the up quark Yukawa matrix 
in Eq.\ref{yuku2}, but we expect that the Cabibbo angle will be reproduced more easily,
with the down quark Yukawa matrix 
in Eq.\ref{yukd2} being the same in both cases.

We also consider a third model with 
the assignments $k_{u^{c}_{3,2,1}}=5,4,-3$
(i.e. differing by the assignment $k_{u^c_2}=4$)
such that the combinations $Qu^{c}_{3}, Qu^{c}_{2}\phi, Qu^{c}_{1}\phi^4$ 
imply the modular forms $Y^{(6)}_{\mathbf{3}}$, $Y^{(6)}_{\mathbf{3}}$, $Y^{(2)}_{\mathbf{3}}$, respectively,
where powers of $\phi$ with $k_{\phi}=1$ are required. Thus with $k_{u^{c}_{3,2,1}}=5,4,-3$ we have:
 \begin{equation}
Y_u^{III} =\begin{pmatrix}
	 ~~&~~     ~~&~~    ~~&~~ \\[-0.1in]
 \alpha_u  \tilde{\phi}^4\,Y_1 ~~&~~ \alpha_u  \tilde{\phi}^4 \,Y_3 ~~&~~ \alpha_u  \tilde{\phi}^4 \,Y_2 \\
 ~~&~~     ~~&~~    ~~&~~ \\[-0.02in]
 \beta^I_u  \tilde{\phi}  Y^{(6)}_{2,I} + \beta^{II}_u  \tilde{\phi}  Y^{(6)}_{2,II}~&~ 
 \beta^I_u  \tilde{\phi}   Y^{(6)}_{1,I}+ \beta^{II}_u   \tilde{\phi}  Y^{(6)}_{1,II}~&~
 \beta^I_u   \tilde{\phi}  Y^{(6)}_{3,I}+\beta^{II}_u  \tilde{\phi}   Y^{(6)}_{3,II}\\
 ~&~        ~&~      ~&~  \\[-0.02in]
 \gamma^I_u  Y^{(6)}_{3,I} + \gamma^{II}_u  Y^{(6)}_{3,II}~&~ 
 \gamma^I_u   Y^{(6)}_{2,I}+ \gamma^{II}_u   Y^{(6)}_{2,II}~&~
 \gamma^I_u   Y^{(6)}_{1,I}+\gamma^{II}_u   Y^{(6)}_{1,II}\\
 ~~&~~     ~~&~~    ~~&~~ 
 \label{yuku4}
 \end{pmatrix} 
 \end{equation}
 
All the above three possibilities for the up quark Yukawa matrices have the third family controlled by a weight $6$ modular form,
resulting from the choice $k_{u^c_3}=5$. We now consider third family modular forms of weight $4$
corresponding to the choice $k_{u^c_3}=3$. This would lead to three more possibilities as shown below. 

With $k_{u^{c}_{3,2,1}}=3,-1,-3$ we have:
\begin{equation}
Y_u^{IV} =\begin{pmatrix}
	 ~~&~~     ~~&~~    ~~&~~ \\[-0.1in]
 \alpha_u  \tilde{\phi}^4\,Y_1 ~~&~~ \alpha_u  \tilde{\phi}^4 \,Y_3 ~~&~~ \alpha_u  \tilde{\phi}^4 \,Y_2 \\
 ~~&~~     ~~&~~    ~~&~~ \\[-0.02in]
\beta_u  \tilde{\phi}^2 Y_2 ~&~ \beta_u  \tilde{\phi}^2 Y_1  ~&~ \beta_u  \tilde{\phi}^2 Y_3 \\
 ~&~        ~&~      ~&~  \\[-0.02in]
 \gamma_u  Y^{(4)}_3 ~&~  \gamma_u Y^{(4)}_2  ~&~ \gamma_u   Y^{(4)}_1 \\
  ~~&~~     ~~&~~    ~~&~~ 
 \label{yuku20}
 \end{pmatrix} 
 \end{equation}
 With $k_{u^{c}_{3,2,1}}=3,2,-3$ we have:
 \begin{equation}
Y_u^{V} =\begin{pmatrix}
	 ~~&~~     ~~&~~    ~~&~~ \\[-0.1in]
 \alpha_u  \tilde{\phi}^4\,Y_1 ~~&~~ \alpha_u  \tilde{\phi}^4 \,Y_3 ~~&~~ \alpha_u  \tilde{\phi}^4 \,Y_2 \\
 ~~&~~     ~~&~~    ~~&~~ \\[-0.02in]
\beta_u  \tilde{\phi} Y^{(4)}_2 ~&~ \beta_u  \tilde{\phi} Y^{(4)}_1  ~&~ \beta_u  \tilde{\phi} Y^{(4)}_3 \\
 ~&~        ~&~      ~&~  \\[-0.02in]
 \gamma_u  Y^{(4)}_3 ~&~  \gamma_u Y^{(4)}_2  ~&~ \gamma_u   Y^{(4)}_1 \\
 ~~&~~     ~~&~~    ~~&~~ 
 \label{yuku30}
 \end{pmatrix} 
 \end{equation}
With $k_{u^{c}_{3,2,1}}=3,4,-3$ we have:
\begin{equation}
Y_u^{VI} =\begin{pmatrix}
	 ~~&~~     ~~&~~    ~~&~~ \\[-0.1in]
 \alpha_u  \tilde{\phi}^4\,Y_1 ~~&~~ \alpha_u  \tilde{\phi}^4 \,Y_3 ~~&~~ \alpha_u  \tilde{\phi}^4 \,Y_2 \\
 ~~&~~     ~~&~~    ~~&~~ \\[-0.02in]
 \beta^I_u  \tilde{\phi}  Y^{(6)}_{2,I} + \beta^{II}_u  \tilde{\phi}  Y^{(6)}_{2,II}~&~ 
 \beta^I_u  \tilde{\phi}   Y^{(6)}_{1,I}+ \beta^{II}_u   \tilde{\phi}  Y^{(6)}_{1,II}~&~
 \beta^I_u   \tilde{\phi}  Y^{(6)}_{3,I}+\beta^{II}_u  \tilde{\phi}   Y^{(6)}_{3,II}\\
 ~&~        ~&~      ~&~  \\[-0.02in]
 \gamma_u  Y^{(4)}_3 ~&~  \gamma_u Y^{(4)}_2  ~&~ \gamma_u   Y^{(4)}_1 \\
  ~~&~~     ~~&~~    ~~&~~ 
 \label{yuku40}
 \end{pmatrix} 
 \end{equation}

 Finally we also consider third family modular forms of weight $2$
corresponding to the choice $k_{u^c_3}=1$. This would lead to three final possibilities as shown below. 

With $k_{u^{c}_{3,2,1}}=1,-1,-3$ we have:
\begin{equation}
Y_u^{VII} =\begin{pmatrix}
	 ~~&~~     ~~&~~    ~~&~~ \\[-0.1in]
 \alpha_u  \tilde{\phi}^4\,Y_1 ~~&~~ \alpha_u  \tilde{\phi}^4 \,Y_3 ~~&~~ \alpha_u  \tilde{\phi}^4 \,Y_2 \\
 ~~&~~     ~~&~~    ~~&~~ \\[-0.02in]
\beta_u  \tilde{\phi}^2 Y_2 ~&~ \beta_u  \tilde{\phi}^2 Y_1  ~&~ \beta_u  \tilde{\phi}^2 Y_3 \\
 ~&~        ~&~      ~&~  \\[-0.02in]
 \gamma_u  Y_3 ~&~  \gamma_u Y_2  ~&~ \gamma_u   Y_1 \\
  ~~&~~     ~~&~~    ~~&~~ 
 \label{yuku200}
 \end{pmatrix} 
 \end{equation}
  With $k_{u^{c}_{3,2,1}}=1,2,-3$ we have:
 \begin{equation}
Y_u^{VIII} =\begin{pmatrix}
	 ~~&~~     ~~&~~    ~~&~~ \\[-0.1in]
 \alpha_u  \tilde{\phi}^4\,Y_1 ~~&~~ \alpha_u  \tilde{\phi}^4 \,Y_3 ~~&~~ \alpha_u  \tilde{\phi}^4 \,Y_2 \\
 ~~&~~     ~~&~~    ~~&~~ \\[-0.02in]
\beta_u  \tilde{\phi} Y^{(4)}_2 ~&~ \beta_u  \tilde{\phi} Y^{(4)}_1  ~&~ \beta_u  \tilde{\phi} Y^{(4)}_3 \\
 ~&~        ~&~      ~&~  \\[-0.02in]
 \gamma_u  Y_3 ~&~  \gamma_u Y_2  ~&~ \gamma_u   Y_1 \\
 ~~&~~     ~~&~~    ~~&~~ 
 \label{yuku300}
 \end{pmatrix} 
 \end{equation}
With $k_{u^{c}_{3,2,1}}=1,4,-3$ we have:
\begin{equation}
Y_u^{IX} =\begin{pmatrix}
	 ~~&~~     ~~&~~    ~~&~~ \\[-0.1in]
 \alpha_u  \tilde{\phi}^4\,Y_1 ~~&~~ \alpha_u  \tilde{\phi}^4 \,Y_3 ~~&~~ \alpha_u  \tilde{\phi}^4 \,Y_2 \\
 ~~&~~     ~~&~~    ~~&~~ \\[-0.02in]
 \beta^I_u  \tilde{\phi}  Y^{(6)}_{2,I} + \beta^{II}_u  \tilde{\phi}  Y^{(6)}_{2,II}~&~ 
 \beta^I_u  \tilde{\phi}   Y^{(6)}_{1,I}+ \beta^{II}_u   \tilde{\phi}  Y^{(6)}_{1,II}~&~
 \beta^I_u   \tilde{\phi}  Y^{(6)}_{3,I}+\beta^{II}_u  \tilde{\phi}   Y^{(6)}_{3,II}\\
 ~&~        ~&~      ~&~  \\[-0.02in]
 \gamma_u  Y_3 ~&~  \gamma_u Y_2  ~&~ \gamma_u   Y_1 \\
  ~~&~~     ~~&~~    ~~&~~ 
 \label{yuku400}
 \end{pmatrix} 
 \end{equation}

 Note that there is only one possibility for the first family
of up quarks since the required suppression $\tilde{\phi}^4$ can only be achieved by modular forms of weight $2$.

In the next section we perform a numerical analysis of our models.
First we check the lepton sector results, based on the matrices in Eqs.\ref{seesawmatrices},\ref{yuke2},
then go on to the quark sector using one of the Yukawa matrices in Eq.\ref{yukd2},\ref{yukd2.2} or \ref{yukd2.3} combined with 
one of Eq.\ref{yuku2}-\ref{yuku400}.  
Without loss of generality we take $ \alpha_{e,d,u} , \beta_{e,d,u},  \gamma_{e,d,u}$ to be real,
with $\beta^{I}_u,\gamma^{I}_u$ real while $\beta^{II}_u,\gamma^{II}_u$ are complex.
We allow $\tilde{\phi}$ to be free but find that the numerical fits prefer $\tilde{\phi} \approx 1/15$, as expected.

\section{Numerical and analytical results}
\label{results}

\subsection{Input data and global analysis}
The charged fermion mass matrices are given by 
\begin{equation}
M_e=Y_e\frac{v_d}{\sqrt{2}}=Y_e\cos \beta \frac{v_H}{\sqrt{2}},\ \ \ \ 
M_d=Y_d\frac{v_d}{\sqrt{2}}=Y_d\cos \beta \frac{v_H}{\sqrt{2}},\ \ \ \ 
M_u=Y_u\frac{v_u}{\sqrt{2}}=Y_u\sin \beta \frac{v_H}{\sqrt{2}}, 
\end{equation}
where the ratio of Higgs VEVs is $\tan \beta = v_u/v_d$ and the SM Higgs VEV is $v_H=\sqrt{v_u^2+v_d^2}=246$ GeV,
and $Y_e,Y_d,Y_u$ represent the Yukawa matrices predicted by the models, namely 
Eq.\ref{yuke2}, for the charged lepton Yukawa matrix, 
Eqs.\ref{yukd2},\ref{yukd2.2} or \ref{yukd2.3} for the down quark Yukawa matrix and 
Eqs.\ref{yuku2}-\ref{yuku400} for the up quark Yukawa matrix.

The scale of Yukawa couplings in this model is given by the string compactification scale, and hence we use couplings calculated at the GUT scale from a minimal SUSY breaking scenario, with $\tan \beta = 5$, as done in \cite{Antusch:2013jca, Bjorkeroth:2015ora,Okada:2019uoy}. Similarly, we use the CKM parameters also at this scale as derived by the same authors. For the charged lepton and down type Yukawa masses, the physical particle masses are given by $m_i ^{MSSM} = y_i ^{MSSM} v_d / \sqrt{2}$, for $i = (e,\mu ,\tau ,d,s,b)$, and for the up quarks, $m_j ^{MSSM} = y_j ^{MSSM} v_u / \sqrt{2}$, for $j=(u,c,t)$. The numerical eigenvalues calculated from our input Yukawa matrices 
$Y_e,Y_d,Y_u$ are matched to $y ^{MSSM}$. Below we list
$\tilde{y}_{i} \equiv y_{i}^{MSSM} \cos \beta $ and $\tilde{y}_{j} \equiv y_{j}^{MSSM} \sin \beta $
for $\tan \beta = 5$, together with the quark mixing parameters,\footnote{These values do not change significantly
for $\tan \beta = 10$. For larger values of $\tan \beta $, threshold corrections become increasingly important.}
\begin{equation}
\begin{alignedat}{5}
\tilde{y}_e&=(1.97\pm 0.0236) \times 10^{-6},~~&&\tilde{y}_\mu&&=(4.16\pm 0.0497) \times 10^{-4},  ~~&&\tilde{y}_\tau&&=(7.07\pm 0.0727) \times 10^{-3}, \\
\rule[15pt]{0pt}{1pt}
\tilde{y}_d&=(4.81\pm 1.06) \times 10^{-6},  &&\tilde{y}_s&&=(9.52\pm 1.03) \times 10^{-5},  &&\tilde{y}_b&&=(6.95\pm 0.175) \times 10^{-3}, \\
\rule[15pt]{0pt}{1pt}
\tilde{y}_u&=(2.92\pm 1.81) \times 10^{-6},  &&\tilde{y}_c&&=(1.43{\pm 0.100}) \times 10^{-3},  &&\tilde{y}_t&&=0.534\pm 0.0341  ~~,\\
\rule[15pt]{0pt}{1pt}
\theta_{12} ^q & =13.027^\circ\pm 0.0814^\circ , &&\theta_{23} ^q&&=2.054^\circ\pm 0.384^\circ, &&\theta_{13} ^q &&=0.1802^\circ\pm 0.0281^\circ~ ,\\
\rule[15pt]{0pt}{1pt}
\delta ^q&=69.21 ^\circ \pm 6.19 ^\circ.
\label{CKMparam}
\end{alignedat}
\end{equation}

For the neutrino parameters, we use the data from NuFit 4.1 (2019) \cite{Esteban:2018azc}, without SK atmospheric data, which we summarise below for Normal Ordering (NO), where we write eorrs in brackets, which correspond to the average of positive and negative $1 \sigma$ deviations. 
\begin{equation}
\begin{alignedat}{5}
\sin^2\theta_{12} &= 0.310(13) ,~~
&&\sin^2\theta_{13}&&  = 0.02241(66) ,~~
&&\sin^2\theta_{23}&&  = 0.558(26) ,\\
\rule[15pt]{0pt}{1pt}
\frac{\Delta m^2_{21}}{10^{-5}~{\rm eV}^2} &= 7.39(21) ,
&&\frac{\Delta m^2_{31}}{10^{-3}~{\rm eV}^2}&& = 2.525(31) ,
&&~~~\delta/\pi&& = 1.23(18)
\end{alignedat}
\end{equation}

For our numerical study, we follow a procedure similar to that described 
in~\cite{Criado:2019tzk}, but here generalised to the 
quark sector, to find the minimum $\chi ^2 _\textrm{min,Q}$ contribution from the CKM and quark Yukawa pulls.
We consider all 27 combinations of $Y_d ^{I} Y_u ^I , \dots , Y_d ^{III} Y_u ^{IX}$, restricting $\tau$ to be within a range which is acceptable to the lepton sector, based on the matrices in Eqs.\ref{seesawmatrices},\ref{yuke2}, 
which is approximately the same as found in~\cite{Ding:2019zxk}, model $\mathcal{D}_{10}$ in their notation. 

\begin{table}
	\centering
	\begin{tabular}{|c|c|c|c|c|c|c|c|c|c|}
		\hline
		&$Y_u ^{I}$ & $Y_u ^{II}$ & $Y_u ^{III}$ & 	$Y_u ^{IV}$ & 	$Y_u ^{V}$ & 	$Y_u ^{VI}$ & 	$Y_u ^{VII}$ & 	$Y_u ^{VIII}$ &	$Y_u ^{IX}$	\rule[-2ex]{0pt}{5ex}\\	\hline		
		$Y_d ^I$ & 118 & 194 & 123 & 420 & 337 & 98.4 & 183 & 244 & 122 \rule[-2ex]{0pt}{5ex} \\
		\hline
		$Y_d ^{II}$ & 79.1 & 78.9 & $\mathbf{0.00}$ & 248 & 279 & 142 & 413 & 479 & 93 \rule[-2ex]{0pt}{5ex} \\
		\hline
	    $Y_d ^{III}$ &  118 & 186 & 135 & 79.8 & 79.5 & $\mathbf{0.00}$ & 117 & 190 & 135 \rule[-2ex]{0pt}{5ex} \\
	    \hline
	\end{tabular}
\caption{$\chi ^2 _\textrm{min,Q}$ for all 27 combinations of $Y_u ,~ Y_d$}
\label{tab:chisq}
\end{table}

We display our best fit points in table \ref{tab:chisq}, for all 27 models. From this table it is clear that 
we find unacceptably high $\chi ^2 _\textrm{min}\gtrsim50$ for all permutations besides $Y_d ^{III} Y_u ^{VI}$ and $Y_d ^{II} Y_u ^{III}$, for which we found an arbitrarily good $\chi ^2 _\textrm{min} <1$. For the remainder of this paper, we will focus on these two successful models, and do not list the benchmark points for the other models which do not well reproduce data.

\subsection{Model $Y_u ^{VI}$, $Y_d ^{III}$}
\subsubsection{Numerical study}

\begin{table}[htp]
	\vspace{-1cm}
	\footnotesize
	\begin{center}
		\noindent\makebox[\textwidth]{
			\begin{tabular}{|c|c |c|  c|c|}
				\cline{1-2} \cline{4-5} 	
				Lepton observable & value (pull) && Quark Observable & value(pull)
				\rule[-2ex]{0pt}{5ex}\\
				\cline{1-2} \cline{4-5}		
				$\Delta m^2_{21} \cdot 10^5~({\rm eV}^2)$ & 7.39 $(0.0)$ && $\theta_{12} ^q$ & 0.227 $(0.0)$
				\rule[-2ex]{0pt}{5ex}\\
				$\Delta m^2_{31} \cdot 10^3~({\rm eV}^{2})$ & 2.52 $(0.0)$ && $\theta_{13} ^q$ & 0.00314 $(0.0)$ 
				\rule[-2ex]{0pt}{5ex}\\
				$\sin ^{2} \theta_{12}$& 0.310 $(0.0)$ && $\theta_{23} ^q$ & 0.0358 $(0.0)$ 
				\rule[-2ex]{0pt}{5ex}\\
				$\sin ^{2} \theta_{13}$& 0.0224 $(0.0)$ && $\delta ^q / \pi$ & 1.21 $(0.0)$ 
				\rule[-2ex]{0pt}{5ex}\\
				$\sin ^{2} \theta_{23}$& 0.562 $(0.2)$ && $y_u\cdot 10^5$ & 1.49 $(0.0)$ 
				\rule[-2ex]{0pt}{5ex}\\	
				$\delta / \pi$ & 1.58 $(1.9)$ &&  $y_c\cdot 10^3$ & 7.29 $(0.0)$ 
				\rule[-2ex]{0pt}{5ex}\\
				$y_e \cdot 10^5$ & 1.00 $(0.0)$ && $y_t$ & 2.72 $(0.0)$ 
				\rule[-2ex]{0pt}{5ex}\\
				$y_\mu \cdot 10^3$ & 2.12 $(0.0)$ && $y_d\cdot 10^5$ & 2.45 $(0.0)$ 
				\rule[-2ex]{0pt}{5ex}\\				
				$y_\tau \cdot 10^2$ & 3.61 $(0.0)$ && $y_s\cdot 10^4$ & 4.85 $(0.0)$ 
				\rule[-2ex]{0pt}{5ex}\\
				$\chi ^{2} _{\textrm{min,L} }$ & 3.67&& $y_b\cdot 10^2$ & 3.54 $(0.0)$ 
				\rule[-2ex]{0pt}{5ex}\\
				\cline{1-2}  	
				Lepton prediction & value && $\chi ^{2} _{\textrm{min,Q} }$ & 0.0
				\rule[-2ex]{0pt}{5ex}\\
				\cline{1-2}  \cline{4-5} 
				$m_1 ~({\rm eV})$&0.11 && Quark input & value 
				\rule[-2ex]{0pt}{5ex}\\
				\cline{4-5}
				$m_2 ~({\rm eV})$ & 0.11&&  $\alpha_u$ & -1.476
				\rule[-2ex]{0pt}{5ex}\\
				$m_3 ~({\rm eV})$ & 0.12  && \begin{tabular}{c}
					$\beta_u ^I$ \rule[-2ex]{0pt}{5ex}\\ $\beta_u ^{II}$
				\end{tabular} &\begin{tabular}{c}~-0.1264 \rule[-2ex]{0pt}{5ex}\\$0.2697-0.1971i$\end{tabular}
				\rule[-2ex]{0pt}{5ex}\\
				$\alpha_{21} /\pi$ & 0.013 && $\gamma_u$ & 2.720
				\rule[-2ex]{0pt}{5ex}\\
				$\alpha_{31} /\pi$ & 1.01  && $\alpha_d$ & -2.387
				\rule[-2ex]{0pt}{5ex}\\
				$m_{ee}~({\rm eV})$ & 0.11  && $\beta_d$ & 2.672
				\rule[-2ex]{0pt}{5ex}\\
				MO & NO && 	$\gamma_d ^I$ & 0.6253
				\rule[-2ex]{0pt}{5ex}\\
				\cline{1-2}  
				Lepton Input & value && $\gamma_d ^{II}$ & $0.4958- 0.2187i$  \rule[-2ex]{0pt}{5ex}\\ 
				\cline{1-2} \cline{4-5}
				Re($g_2/g_1$) & 0.4185 &\multicolumn{2}{c}{} \rule[-2ex]{0pt}{5ex} \\
				\cline{4-5}
				Im($g_2/g_1$) &1.048 &&Common Input&value \rule[-2ex]{0pt}{5ex} \\
				\cline{4-5}
				$g_1 ^2 v_u ^2 /\Lambda$ (eV)&0.05506 &&Re($\tau$) & 0.03610\rule[-2ex]{0pt}{5ex}\\
				$\alpha_e$ & -0.9778 &&Im($\tau$) & 2.352 \rule[-2ex]{0pt}{5ex}\\
				$\beta_e$ & -0.6615 &&  $\tilde{\phi}$ & 0.05663 \rule[-2ex]{0pt}{5ex}\\
				\cline{4-5}
					$\gamma_e$ & -0.6360 \rule[-2ex]{0pt}{5ex}\\
				\cline{1-2}
		\end{tabular}}
		\caption{Results of the fit to lepton and quark data for model combining $M_\nu ,~ Y_e ,~ Y_u ^{VI} ,~Y_d ^{III}$. In the left panel are the lepton observables and pulls (in fractions of $1 \sigma$), the $\chi ^2 _\textrm{min,L}$ contribution from the lepton sector, as well as predictions for neutrino masses, phases, neutrinoless double beta decay and mass ordering. The inputs for the lepton sector are displayed at the bottom. In the right panel we have the quark observables and pulls, the $\chi^2 _\textrm{min,Q}$ quark contribution, and quark inputs. At the bottom right we list the $\tau$ and $\phi$ inputs which are common to both sectors. We note that $\tilde{\phi}=1/15=0.06667$ for example may be fixed exactly to find an equivalently good benchmark point.}
		\label{tab:fit1}
	\end{center}
\end{table}	

We find two combinations of down and up quark Yukawa matrices has an acceptable $\chi^2 _\textrm{min,Q}$ value, from $Y_u ^{VI}$ in combination with $Y_d ^{III}$, which we study in this section and $Y_u ^{III},~Y_d ^{II}$ which we study in the next \footnote{We have also tested these models with a different $\tan \beta = 10$, to check we are not overly sensitive to this initial choice.} In Tab. \ref{tab:fit1} we write the input and output parameters, both for the quark and lepton sectors for our best fit point in this model. Since the neutrino sector is the same as found in \cite{Ding:2019zxk}, model $\mathcal{D}_{10}$, and charged lepton Yukawa matrix a similar form besides the addition of weightons, the lepton observables and predictions are similar to what is seen by Ding, et. al. However, the quark sector is entirely new of our own construction. Here we see that by tuning the $\alpha_{i} ,~ \beta_i ,~ \gamma_i$ parameters to match SM fermion Yukawa couplings (at GUT scale), we also find very strong agreement with the CKM angles and phase. 

To explain why this is the case, we first look at numerical motivations and then go on to study the analytic properties of this point. Firstly, we list the two numerical mixing matrices which produce the CKM for our best fit point found in Tab.  \ref{tab:fit1} are as follows. Defining ${U_L ^{u,VI}} ^\dagger ({Y_u^{VI}}^{\dagger} Y_u^{VI} ) U_L ^{u,VI} = {Y_u^{VI}} ^{diag}$, and similarly for the down sector we find the following two diagonalising matrices,
\begin{align}
U_L ^{u,VI} &= 
\left(
\begin{array}{ccc}
0.981 & -0.193 & -0.00283 \\
-0.149+0.122 i & -0.758+0.622 i & -0.0433-0.00328 i \\
0.00411-0.00522 i & 0.0355-0.0244 i & -0.988-0.15 i \\
\end{array}
\right),\\
U_L ^{d,III} &= 
\left(
\begin{array}{ccc}
-0.999 & 0.0436 & -0.00277 \\
-0.0434-0.00329 i & -0.996-0.0736 i & 0.00225+0.0273 i \\
-0.0028-0.00043 i & -0.00319-0.0272 i & 0.962-0.273 i \\
\end{array}
\right).
\end{align}

We can see that the Cabbibo angle, $\theta_{12}$ is mostly generated by the mixing in the up sector, since ${U_L ^u}^{1,2} \gg {U_L ^d}^{1,2}$, however both sectors will play a similar role in generating the other two angles,
with the imaginary part playing a significant role for $\theta_{23}$. To examine this further, we turn to an analytic study of the structure of our up and down Yukawa matrices.		

\subsubsection{Analytic results}

We may approximate the analytic forms of our successful model of $Y_u ^{VI}$, $Y_d ^{III}$ rewriting the weighton and weight two modular forms as follows, using Eq.\ref{Y2},
and writing the higher weight forms directly in terms of these weight two approximations as in Eqs. \ref{weight4}, \ref{weight6},
\begin{equation}
\tilde{\phi} \simeq 0.057 \equiv \epsilon_1  ,~~~~ 
\end{equation}
\begin{equation}
\begin{pmatrix}Y_1(\tau)\\Y_2(\tau)\\Y_3(\tau)\end{pmatrix} 
=
\begin{pmatrix}1 + \mathcal{O} (q) \\-6q ^{1/3} + \mathcal{O} (q) \\ -18 q ^{2/3} + \mathcal{O} (q) \end{pmatrix}
\simeq
\begin{pmatrix}1.00\\-0.043 - 0.0033i\\-0.00094-0.00014i\end{pmatrix}
\equiv \begin{pmatrix}1\\ \epsilon_2 \\ \epsilon_3\end{pmatrix}   .
\label{exp1}
\end{equation}
where $q=e^{2\pi i\tau}$.
We find numerically that $\epsilon_1 ^2$ is a similar order to $\epsilon_2 ^2$, and to ${\epsilon_3}$. Consequently, we may take the first non-trivial term at the order $\mathcal{O} (\epsilon_i ) \sim \mathcal{O} (\epsilon_1 ) \sim \mathcal{O} (\epsilon_2 ) \sim \mathcal{O} (\epsilon_3 ^{1/2})$, dropping higher corrections in each entry of our successful model. We find the following results for the up and down quark Yukawa matrices, respectively,
making a leading order approximation for each element of the matrix,
\begin{equation}
\hspace{-2cm}
Y_u^{VI} \simeq
\begin{pmatrix}
~~&~~     ~~&~~    ~~&~~ \\[-0.1in]
\epsilon _1 ^4 \alpha _u & \epsilon _1^4 \epsilon _3 \alpha _u & \epsilon _1^4 \epsilon _2 \alpha _u \\
~~&~~     ~~&~~    ~~&~~ \\[-0.1in]
\epsilon _1 \epsilon _2 \left(2 \beta _u^{{II}}+\beta _u^{I} \right) & \epsilon _1 \beta _u ^{I} & \epsilon _1 \left(2 \epsilon _2^2 \beta _u^{{II}}+\epsilon _3 \beta _u ^{I} \right) \\
~~&~~     ~~&~~    ~~&~~ \\[-0.1in]
\left(\epsilon _2^2-\epsilon _3\right) \gamma _u & - \epsilon _2 \gamma _u & \gamma _u \\
~~&~~     ~~&~~    ~~&~~ \\[-0.1in]
\end{pmatrix},
\end{equation}

\begin{equation}
\hspace{-2cm}
Y_d^{III} \simeq
\begin{pmatrix}
~~&~~     ~~&~~    ~~&~~ \\[-0.1in]
\epsilon _1^4 \alpha _d & \epsilon _1^4 \epsilon _3 \alpha _d & \epsilon _1^4 \epsilon _2 \alpha _d \\
~~&~~     ~~&~~    ~~&~~ \\[-0.1in]
\epsilon _1^3 \epsilon _2 \beta _d & \epsilon _1^3 \beta _d & \epsilon _1^3 \epsilon _3 \beta _d \\
~~&~~     ~~&~~    ~~&~~ \\[-0.1in]
\epsilon _1 \left(2 \epsilon _2^2 \gamma _d^{{II}}+\epsilon _3 \gamma _d^I\right) & \epsilon _1 \epsilon _2 \left(2 \gamma _d^{{II}}+\gamma _d^I\right) & \epsilon _1 \gamma _d^I \\
~~&~~     ~~&~~    ~~&~~ \\[-0.1in]
\end{pmatrix}.
\end{equation}

Since the matrices are hierarchical, we can make an estimate for the three mixing angles as follows, which accurately reproduces the fully calculated CKM angles. We then express $\epsilon_2,\epsilon_3$ by using the q-expansions, for which the first order reproduces well the full Dedekind-eta value for our best fit point.
\begin{alignat}{2}
\theta_{12}&\simeq \left| \frac{Y_u ^{2,1}}{Y_u ^{2,2}} - \frac{Y_d ^{2,1}}{Y_d ^{2,2}} \right| =\left| 2\epsilon_2 \frac{\beta _u ^{{II}}}{\beta _u ^{{I}}} \right|
&&=12 e^{-\frac{2}{3} \pi  \Im(\tau )} \left| \frac{\beta _u^{\text{II}}}{\beta _u^{\text{I}}} \right| \label{12}\\
\theta_{13}&\simeq \left| \frac{Y_u ^{3,1}}{Y_u ^{3,3}} - \frac{Y_d ^{3,1}}{Y_d ^{3,3}} \right| =\left|\epsilon_2 ^2\frac{\gamma_d ^{I} - 2 \gamma _d ^{{II}}}{\gamma _d ^{I}} -2 \epsilon_3 \right|
&&=72 e^{-\frac{4}{3} \pi  \Im(\tau )} \left| 1-\frac{{\gamma _d ^{II}}}{{\gamma _d ^I
	}}\right| \label{13}\\
\theta_{23}&\simeq \left| \frac{Y_u ^{3,2}}{Y_u ^{3,3}} - \frac{Y_d ^{3,2}}{Y_d ^{3,3}} \right| =\left|2 \epsilon_2 \frac{\gamma_d ^{I} + \gamma _d ^{{II}}}{\gamma _d ^{I}}  \right|
&&=12 e^{-\frac{2}{3} \pi  \Im(\tau )} \left| 1+\frac{{\gamma _d ^{II}}}{{\gamma _d ^I
	}}\right|
	\label{23}
\end{alignat}
The above approximations reproduce the numerical values of the quark mixing angles well, to two significant figures for $\theta_{12},~\theta_{23}$, but only within a factor of two for $\theta_{13}$. This is because it is the smallest angle, and hence sensitive to additional contributions. For the two larger angles, there are several reasons why the above expressions well reproduce data. To begin with, quark mixing angles are all small, so a small angle approximation is valid. Furthermore, overall factors and phases cancel in the ratios such as 
$ \frac{Y_u ^{2,1}}{Y_u ^{2,2}}$ and $\frac{Y_d ^{2,1}}{Y_d ^{2,2}}$, since each row of the Yukawa matrices is 
controlled by a particular modular form, therefore the physical CKM angles are identified as the difference in these two ratios, with no arbitrary relative phase. This is quite different from a traditional FN model based on an Abelian symmetry, where mixing angle predictions would depend on arbitrary coefficients and phases.
It implies that partial cancellations occur between $ \frac{Y_u ^{2,1}}{Y_u ^{2,2}}$ and $\frac{Y_d ^{2,1}}{Y_d ^{2,2}}$
in constructing $\theta_{12}$, which leads to a particularly simple form without $\beta _u ^I$ in the numerator.
It also implies that the mixing angles are independent of $\epsilon_1$ which cancels 
in the ratios, so the only role of $\epsilon_1$ is to control mass hierarchies.
The mixing angles are therefore completely controlled by $\epsilon_{2}$ and $\epsilon_{3}$,
which however are not independent parameters, being related by the 
expansion of the $A_4$ triplet modular forms in Eq.\ref{Y2}. 
This dependence is manifested in the final expressions on the 
right-hand sides of the Eqs.\ref{12},\ref{13},\ref{23} based on the truncations in Eq.\ref{exp1},
which are valid for small $q=e^{2\pi i\tau}$ when the imaginary part of $\tau$ is large.
Despite the large prefactors, the CKM angles are therefore 
small due to an exponential suppression arising from the 
best fit point $\tau$ having a large imaginary part. 
One can see that in the limit $\tau \rightarrow i \infty$ the CKM angles go to zero, which is expected as this would correspond to diagonal Yukawa matrices. Given $\mathcal{O}(1)$ input parameters, we then see the required value of $\tau$ to match the observed CKM values must be near $\tau \simeq 2.35 i$.

\subsection{Model $Y_u ^{III}$, $Y_d ^{II}$}
\subsubsection{Numerical study}
We now study a second successful model, comprised of $Y_u ^{III}$, $Y_d ^{II}$, with input and output parameters found in Tab. \ref{tab:fit2}. This section will proceed analagously to the previous one.

\begin{table}[htp]
	\vspace{-1cm}
	\footnotesize
	\begin{center}
		\noindent\makebox[\textwidth]{
			\begin{tabular}{|c|c |c|  c|c|}
				\cline{1-2} \cline{4-5} 	
				Lepton observable & value (pull) && Quark Observable & value(pull)
				\rule[-2ex]{0pt}{5ex}\\
				\cline{1-2} \cline{4-5}		
				$\Delta m^2_{21} \cdot 10^5~({\rm eV}^{2})$ & 7.39 $(0.0)$ && $\theta_{12} ^q$ & 0.227 $(0.0)$
				\rule[-2ex]{0pt}{5ex}\\
				$\Delta m^2_{31} \cdot 10^3~({\rm eV}^{2})$ & 2.52 $(0.0)$ && $\theta_{13} ^q$ & 0.00314 $(0.0)$ 
				\rule[-2ex]{0pt}{5ex}\\
				$\sin ^{2} \theta_{12}$& 0.310 $(0.0)$ && $\theta_{23} ^q$ & 0.0358 $(0.0)$ 
				\rule[-2ex]{0pt}{5ex}\\
				$\sin ^{2} \theta_{13}$& 0.0224 $(0.0)$ && $\delta ^q / \pi$ & 1.21 $(0.0)$ 
				\rule[-2ex]{0pt}{5ex}\\
				$\sin ^{2} \theta_{23}$& 0.562 $(0.2)$ && $y_u\cdot 10^5$ & 1.49 $(0.0)$ 
				\rule[-2ex]{0pt}{5ex}\\	
				$\delta / \pi$ & 1.58 $(1.9)$ &&  $y_c\cdot 10^3$ & 7.29 $(0.0)$ 
				\rule[-2ex]{0pt}{5ex}\\
				$y_e \cdot 10^5$ & 1.00 $(0.0)$ && $y_t$ & 2.72 $(0.0)$ 
				\rule[-2ex]{0pt}{5ex}\\
				$y_\mu \cdot 10^3$ & 2.12 $(0.0)$ && $y_d\cdot 10^5$ & 2.45 $(0.0)$ 
				\rule[-2ex]{0pt}{5ex}\\				
				$y_\tau \cdot 10^2$ & 3.61 $(0.0)$ && $y_s\cdot 10^4$ & 4.85 $(0.0)$ 
				\rule[-2ex]{0pt}{5ex}\\
				$\chi ^{2} _{\textrm{min,L} }$ & 3.67&& $y_b\cdot 10^2$ & 3.54 $(0.0)$ 
				\rule[-2ex]{0pt}{5ex}\\
				\cline{1-2}  	
				Lepton prediction & value && $\chi ^{2} _{\textrm{min,Q} }$ & 0.0
				\rule[-2ex]{0pt}{5ex}\\
				\cline{1-2}  \cline{4-5} 
				$m_1 ~({\rm eV})$&0.11 && Quark input & value 
				\rule[-2ex]{0pt}{5ex}\\
				\cline{4-5}
				$m_2 ~({\rm eV})$ & 0.11&&  $\alpha_u$ & -1.137
				\rule[-2ex]{0pt}{5ex}\\
				$m_3 ~({\rm eV})$ & 0.12  && \begin{tabular}{c}
					$\beta_u ^I$ \rule[-2ex]{0pt}{5ex}\\ $\beta_u ^{II}$
				\end{tabular} &\begin{tabular}{c}~-0.1048 \rule[-2ex]{0pt}{5ex}\\$0.1937+0.1985i$\end{tabular}
				\rule[-2ex]{0pt}{5ex}\\
				$\alpha_{21} /\pi$ & 0.012 && \begin{tabular}{c}
					$\gamma_u ^I$ \rule[-2ex]{0pt}{5ex}\\ $\gamma_u ^{II}$
				\end{tabular} &\begin{tabular}{c}~2.722 \rule[-2ex]{0pt}{5ex}\\$-1.697-0.4260i$\end{tabular}
				\rule[-2ex]{0pt}{5ex}\\
				$\alpha_{31} /\pi$ & 1.01  && $\alpha_d$ & -1.137
				\rule[-2ex]{0pt}{5ex}\\
				$m_{ee} ~({\rm eV})$ & 0.11  && $\beta_d$ & -1.533
				\rule[-2ex]{0pt}{5ex}\\
				MO & NO && 	$\gamma_d$ & -0.5194
				\rule[-2ex]{0pt}{5ex}\\
				\cline{1-2} \cline{4-5} 
				Lepton Input & value & \multicolumn{2}{c}{}   \rule[-2ex]{0pt}{5ex}\\ 
				\cline{1-2} 
				Re($g_2/g_1$) & 0.4185 &\multicolumn{2}{c}{} \rule[-2ex]{0pt}{5ex} \\
				\cline{4-5}
				Im($g_2/g_1$) &1.038 &&Common Input&value \rule[-2ex]{0pt}{5ex} \\
				\cline{4-5}
				$g_1 ^2 v_u ^2 /\Lambda$ (eV${}$)&0.05508 &&Re($\tau$) & 0.03610\rule[-2ex]{0pt}{5ex}\\
				$\alpha_e$ & -0.4658 &&Im($\tau$) & 2.353 \rule[-2ex]{0pt}{5ex}\\
				$\beta_e$ & -0.4566 &&  $\tilde{\phi}$ & -0.06816 \rule[-2ex]{0pt}{5ex}\\
				\cline{4-5}
				$\gamma_e$ & 0.5284 \rule[-2ex]{0pt}{5ex}\\
				\cline{1-2}
		\end{tabular}}
		\caption{Results of the fit to lepton and quark data for model combining $M_\nu ,~ Y_e ,~ Y_u ^{III} ,~Y_d ^{II}$. In the left panel are the lepton observables and pulls (in fractions of $1 \sigma$), the $\chi ^2 _\textrm{min,L}$ contribution from the lepton sector, as well as predictions for neutrino masses, phases, neutrinoless double beta decay and mass ordering. The inputs for the lepton sector are displayed at the bottom. In the right panel we have the quark observables and pulls, the $\chi^2 _\textrm{min,Q}$ quark contribution, and quark inputs. At the bottom right we list the $\tau$ and $\phi$ inputs which are common to both sectors. We note that $\tilde{\phi}=1/15=0.06667$ for example may be fixed exactly to find an equivalently good benchmark point.}
		\label{tab:fit2}
	\end{center}
\end{table}	

We first present the two numerical diagonalising matrices as before. The numerical values of this model are similar to the previous scenario, where $\theta_{12}$ is dominated by the contribution from the up quark sector.

\begin{align}
U_L ^{u,III} &= 
\left(
\begin{array}{ccc}
-0.98-0.0000121 i & 0.198+0.000225 i & 0.00312-0.00123 i \\
0.102+0.17 i & 0.506+0.84 i & -0.00896+0.0148 i \\
-0.00166-0.00417 i & 0.00782-0.0152 i & -0.999+0.047 i \\
\end{array}
\right),\\
U_L ^{d,II} &= 
\left(
\begin{array}{ccc}
0.999-0.000038 i & 0.0437+0.000215 i & -0.00275+0.00068 i \\
0.0434+0.00329 i & -0.995-0.0803 i & -0.0428+0.00722 i \\
-0.00467-0.000712 i & 0.0428+0.00673 i & -0.995+0.091 i \\
\end{array}
\right).
\end{align}

\subsubsection{Analytic results}

We again proceed with the same analytic approach as before, and will find very similar analytic approximations as with the previous scenario. For the new scenario (with slightly different input values of $(\varphi ,~\tau)$), we again see the relation $\mathcal{O} (\epsilon_i ) \sim \mathcal{O} (\epsilon_1 ) \sim \mathcal{O} (\epsilon_2 ) \sim \mathcal{O} (\epsilon_3 ^{1/2})$, and take the lowest non trivial order in each entry in the two Yukawa matrices.
\begin{equation}
\tilde{\phi} \simeq -0.068 \equiv \epsilon_1  ,~~~~ 
\end{equation}
\begin{equation}
\begin{pmatrix}Y_1(\tau)\\Y_2(\tau)\\Y_3(\tau)\end{pmatrix} 
=
\begin{pmatrix}1 + \mathcal{O} (q) \\-6q ^{1/3} + \mathcal{O} (q) \\ -18 q ^{2/3} + \mathcal{O} (q) \end{pmatrix}
\simeq
\begin{pmatrix}1.00\\-0.043 - 0.0033i\\-0.00093-0.00014i\end{pmatrix}
\equiv \begin{pmatrix}1\\ \epsilon_2 \\ \epsilon_3\end{pmatrix}   .
\end{equation}

\begin{equation}
\hspace{-0cm}
Y_u^{III} \simeq
\begin{pmatrix}
~~&~~     ~~&~~    ~~&~~ \\[-0.1in]
\epsilon _1 ^4 \alpha _u & \epsilon _1^4 \epsilon _3 \alpha _u & \epsilon _1^4 \epsilon _2 \alpha _u \\
~~&~~     ~~&~~    ~~&~~ \\[-0.1in]
\epsilon _1 \epsilon _2 \left(2 \beta _u^{{II}}+\beta _u^{I} \right) & \epsilon _1 \beta _u ^{I} & \epsilon _1 \left(2 \epsilon _2^2 \beta _u^{{II}}+\epsilon _3 \beta _u ^{I} \right) \\
~~&~~     ~~&~~    ~~&~~ \\[-0.1in]
\epsilon_3 \gamma_u ^I + 2 \epsilon_2 ^2 \gamma_u ^{II} & \epsilon _2 \left( \gamma_u ^I + 2 \gamma_u ^{II} \right) & \gamma _u ^I \\
~~&~~     ~~&~~    ~~&~~ \\[-0.1in]
\end{pmatrix}
\end{equation}

\begin{equation}
\hspace{-2cm}
Y_d^{II} \simeq
\begin{pmatrix}
~~&~~     ~~&~~    ~~&~~ \\[-0.1in]
\epsilon _1^4 \alpha _d & \epsilon _1^4 \epsilon _3 \alpha _d & \epsilon _1^4 \epsilon _2 \alpha _d \\
~~&~~     ~~&~~    ~~&~~ \\[-0.1in]
\epsilon _1^3 \epsilon _2 \beta _d & \epsilon _1^3 \beta _d & \epsilon _1^3 \epsilon _3 \beta _d \\
~~&~~     ~~&~~    ~~&~~ \\[-0.1in]
\epsilon _1 \gamma_d \left(\epsilon _2^2 - \epsilon_3\right) & -\epsilon _1 \epsilon _2 \gamma _d & \epsilon _1 \gamma _d \\
~~&~~     ~~&~~    ~~&~~ \\[-0.1in]
\end{pmatrix}.
\end{equation}

We now follow the same procedure to approximate the three CKM mixing angles, and replace $\epsilon _{2,3}$ with the q-expansions to first order.
\color{black}
\begin{alignat}{2}
\theta_{12}&\simeq \left| \frac{Y_u ^{2,1}}{Y_u ^{2,2}} - \frac{Y_d ^{2,1}}{Y_d ^{2,2}} \right| =\left| 2\epsilon_2 \frac{\beta _u ^{{II}}}{\beta _u ^{{I}}} \right|
&&=12 e^{-\frac{2}{3} \pi  \Im(\tau )} \left| \frac{\beta _u^{\text{II}}}{\beta _u^{\text{I}}} \right| \\
\theta_{13}&\simeq \left| \frac{Y_u ^{3,1}}{Y_u ^{3,3}} - \frac{Y_d ^{3,1}}{Y_d ^{3,3}} \right| =\left|\epsilon_2 ^2\frac{\gamma_u ^{I} - 2 \gamma _u ^{{II}}}{\gamma _u ^{I}} -2 \epsilon_3 \right|
&&=72 e^{-\frac{4}{3} \pi  \Im(\tau )} \left| 1-\frac{{\gamma _u ^{II}}}{{\gamma _u ^I
}}\right| \\
\theta_{23}&\simeq \left| \frac{Y_u ^{3,2}}{Y_u ^{3,3}} - \frac{Y_d ^{3,2}}{Y_d ^{3,3}} \right| =\left|2 \epsilon_2 \frac{\gamma_u ^{I} + \gamma _u ^{{II}}}{\gamma _u ^{I}}  \right|
&&=12 e^{-\frac{2}{3} \pi  \Im(\tau )} 
 \left| 1+\frac{{\gamma _u ^{II}}}{{\gamma _u ^I
}}\right| 
\end{alignat}

The analytic forms here are identical to the previous model, exchanging $\gamma _d$ previously seen with $\gamma _u$ here. In this scenario, the weight six entries previously found in the third row of $Y_d ^{VI}$ are instead found in the third row of $Y_u ^{III}$. In this scenario, the mixing angles are even more controlled by the up sector than beore.

It can now be understood that these two specific models are both successful, as they both predict the same expressions for the CKM mixing angles above, for which values of $\alpha_i ,~\beta_i ,~\gamma _i$ that explain well the Yukawa couplings of the quarks also well reproduce the observed mixings in the quark sector.

\subsection{Analytic expansion of the lepton matrices}
Finally, it is interesting to apply the same analytic expansion procedure used for the quarks, also to the leptons.
For the charged lepton Yukawa matrix in Eq.\ref{yuke2}, we find (without dropping any terms since the leading order matrix arises at weight 2),
\begin{equation}
Y_e \simeq \begin{pmatrix}
~~&~~     ~~&~~    ~~&~~ \\[-0.1in]
\alpha_e  \epsilon _1 ^4 ~~&~~ \alpha_e  \epsilon _1 ^4 \,\epsilon_3 ~~&~~ \alpha_e  \epsilon _1 ^4 \,\epsilon_2 \\
~~&~~     ~~&~~    ~~&~~ \\[-0.02in]
\beta_e  \epsilon _1 ^2 \epsilon_2 ~&~ \beta_e  \epsilon _1 ^2   ~&~ \beta_e  \epsilon _1 ^2 \epsilon_3 \\
~&~        ~&~      ~&~  \\[-0.02in]
\gamma_e  \epsilon _1  \epsilon_3~&~ \gamma_e  \epsilon _1  \epsilon_2~&~\gamma_e  \epsilon _1  \\
~~&~~     ~~&~~    ~~&~~ 
\end{pmatrix} .
\end{equation}
This structure provides a natural explanation of the charged lepton mass hierarchy, namely 
$m_e:m_{\mu}:m_{\tau}= \alpha_e  \epsilon _1 ^4:\beta_e  \epsilon _1 ^2:\gamma_e  \epsilon _1 $.

After the seesaw mechanism, by inputting and expanding the matrices in Eq.\ref{seesawmatrices}
we find the effective neutrino mass matrix,
\begin{equation}
M_\nu \simeq
g_1 ^2  \frac{v_u ^2}{ \Lambda}
\begin{pmatrix}
~~&~~     ~~&~~    ~~&~~ \\[-0.1in]
-2 & \epsilon _3 & \epsilon _2 \\
~~&~~     ~~&~~    ~~&~~ \\[-0.1in]
\epsilon _3  & -2\epsilon _2 & 1 \\
~~&~~     ~~&~~    ~~&~~ \\[-0.1in]
\epsilon_2 & 1 & -2 \epsilon_3 \\
~~&~~     ~~&~~    ~~&~~ \\[-0.1in]
\end{pmatrix}
+
g_2 ^2 \frac{v_u ^2}{ \Lambda}
\begin{pmatrix}
~~&~~     ~~&~~    ~~&~~ \\[-0.1in]
0 & -2 \epsilon_2 ^2 + \epsilon _3 & \epsilon _2 \\
~~&~~     ~~&~~    ~~&~~ \\[-0.1in]
-2 \epsilon_2 ^2  & 2\epsilon _2 & -1 \\
~~&~~     ~~&~~    ~~&~~ \\[-0.1in]
\epsilon_2 & -1 & -2 \epsilon_2 ^2 + 2 \epsilon_3 \\
~~&~~     ~~&~~    ~~&~~ \\[-0.1in]
\end{pmatrix}.
\end{equation}
The parameters $g_1,g_2$ and $ \epsilon_2,  \epsilon_3$
are determined by the fit to the neutrino mass squared differences and PMNS mixing parameters, which arise predominantly from the neutrino sector, due to the very small charged lepton mixing corrections. 
The large elements in the neutrino mass matrix occurring in the $(1,1)$ and $(2,3)$ positions, controlled by $g_1,g_2$, are responsible for the quasi-degenerate neutrino masses $m_1\sim m_2\sim 0.11$ eV,
and $m_3\sim 0.12$ eV,
with the neutrinoless double beta becay parameter $m_{ee}\sim 0.11$ eV in the sensitivity region of current experiments,
and the cosmological sum of neutrino masses $\sum m_i\sim 0.34$ eV being in the disfavoured region.
Either this lepton model will be discovered soon or it will be excluded in the near future.
In any case we remark that the neutrino sector considered here is identical to that of the Feruglio model, being independent
of the weighton $\phi$, and hence $ \epsilon_1$.

\section{Conclusion}
\label{conclusion}

In this paper we have shown how quark and lepton mass hierarchies can be reproduced in the framework of modular symmetry.
The mechanism we have proposed is analogous to the Froggatt-Nielsen (FN) mechanism, but without requiring any Abelian symmetry 
to be introduced, nor any Standard Model (SM) singlet flavon to break it. The modular weights of fermion fields 
play the role of FN charges, and SM singlet fields with non-zero modular weight called weightons play the role of flavons.

We have illustrated the mechanism by analysing $A_4$ (modular level 3) models of quark and lepton (including neutrino) masses and mixing, with a single modulus field. 
We showed how a previously proposed $A_4$ modular model of leptons can be recast in natural form by introducing a single weighton, then applied similar ideas to 27 possible models in the quark sector. 
We analysed all the quark models, combined with the natural lepton model, and 
identified two viable combinations,
which can successfully describe all quark and lepton (including neutrino) masses and mixing, using a single modulus field $\tau$, and in which all charged fermion mass hierarchies originate from a single weighton. 

We have discussed these two particular examples in some detail, both numerically and analytically, showing how both fermion mass and mixing hierarchies emerge from the modular symmetry. The analytic results clearly show how the fermion mass hierarchies are controlled by the powers of the weighton field which multiply a particular row of the Yukawa matrix, while the smallness of the quark mixing angles arises because of the proximity of the 
modulus field to the fixed point case $\tau_T=i\infty$, which results in exponentially suppressed entries within
a particular row of the Yukawa matrix. This leads to a simple analytic understanding of the smallness of quark mixing angles.

We emphasise that the mechanism introduced in this paper 
is quite unlike the traditional FN mechanism, based on an Abelian symmetry, in which the suppression of both rows and columns of the Yukawa matrices arises from FN charges. In the present approach, fermion mass hierarchies and small quark mixing angles emerge from different aspects of the modular symmetry, without having to introduce an extra Abelian symmetry and an additional flavon to break it. The $A_4$ flavour symmetry arises as a finite subgroup of the underlying modular symmetry, and the weightons responsible for the charged fermion mass hierarchies are $A_4$ singlets which do not break the flavour symmetry.

Finally we note that the class of modular level 3 (with even weight modular forms) examples
of the mechanism we present here are by no means exhaustive.
The new mechanism may be applied to other levels and choices of weights, and to models with any number of moduli fields and weightons.

\vspace{-0.5cm}

\subsection*{Acknowledgements}

S.\,F.\,K. acknowledges the STFC Consolidated Grant ST/L000296/1 and the European Union's Horizon 2020 Research and Innovation programme under Marie Sk\l{}odowska-Curie grant agreements Elusives ITN No.\ 674896 and InvisiblesPlus RISE No.\ 690575.

\end{document}